\documentclass[a4paper,11pt]{article}
\usepackage{jcappub}
\usepackage{lineno}
\usepackage{graphicx}
\usepackage[normalem]{ulem}
\usepackage{float}
\usepackage{amsmath}
\usepackage{amssymb}
\usepackage{booktabs}
\usepackage{hhline}
\usepackage{multirow}
\usepackage[rightcaption]{sidecap}
\usepackage{hyperref}
\usepackage{xcolor}
\usepackage{wrapfig}
\usepackage{braket}
\usepackage{comment}
\usepackage{orcidlink}
\graphicspath{{./Figures/}}

\renewcommand{\d}{\mathrm{d}}
\def\beq{\begin{equation}}
\def\eeq{\end{equation}}
\def\bea{\begin{eqnarray}}
\def\eea{\end{eqnarray}}
\def\l{\left}
\def\r{\right}

\let\vec\mathbf

\title{\boldmath Scalar-Induced Gravitational Waves Probing Phase Transitions in the Standard Model and Beyond}

\author[a]{Peizhi Du\,\orcidlink{0000-0001-6801-4829},}
\author[b,c]{Xiang-Xi Zeng\,\orcidlink{0009-0003-0632-3230},}
\author[d]{Jiahang Zhong\,\orcidlink{0009-0000-6304-0681}}

\emailAdd{dupeizhi@ustc.edu.cn}
\emailAdd{zengxiangxi@sjtu.edu.cn}
\emailAdd{jzhongay@connect.ust.hk}

\affiliation[a]{Laboratory of Spin Magnetic Resonance, School of Physical Sciences, Anhui Province Key Laboratory of Scientific Instrument Development and Application, University of Science and Technology of China, Hefei 230026, China}

\affiliation[b]{State Key Laboratory of Dark Matter Physics, Tsung-Dao Lee Institute and School of Physics and Astronomy, Shanghai Jiao Tong University, Shanghai 201210, China}

\affiliation[c]{Key Laboratory for Particle Astrophysics and Cosmology (MOE), and Shanghai Key Laboratory for Particle Physics and Cosmology, Shanghai Jiao Tong University, Shanghai 201210, China}

\affiliation[d]{Department of Physics and Jockey Club Institute for Advanced Study,
The Hong Kong University of Science and Technology, Hong Kong S.A.R., China}

\abstract{Scalar-induced gravitational waves (SIGWs) are unique messengers of physics in the early Universe and possible sources for the observed data from pulsar timing arrays (PTAs). It has been demonstrated that SIGWs have the potential to probe cosmological phase transitions via measurement of the equation of state and sound speed of the cosmic thermal bath, which have non-trivial time dependence during phase transitions. In this work, we improve on existing studies of SIGWs in a time-dependent background with time-dependent initial conditions and an entropy-conserving thermal history of the Standard Model (SM). We provide possible physical explanations for features of the gravitational wave spectrum from phase transitions, the most prominent of which for a scale-invariant curvature spectrum is a step due to entropy dilution. 
We show that current and future gravitational wave observations across a wide range of frequencies can probe phase transitions in the SM and physics beyond the Standard Model (BSM). Signals from the QCD crossover appear around 10 nHz, inside the observable window of PTAs. We perform data analyses using NANOGrav 15-year data, showing a mild preference for the QCD crossover over standard radiation domination for two choices of curvature spectra. BSM phase transitions in the TeV range lead to signals around mHz, which are accessible in future space-based experiments such as LISA, Taiji and TianQin.}

\begin{document}

\maketitle

\section{Introduction}

Gravitational waves (GWs) are a powerful probe of the primordial Universe. Because they
free-stream through the cosmic plasma, GWs can access information from much earlier times
than the CMB photons, 
to which the era prior to
recombination is opaque. This prospect drives a broad experimental program. Ground-based detectors such as
LIGO/Virgo/KAGRA operate at kilohertz frequencies~\cite{LIGOScientific:2016aoc,KAGRA:2021vkt,KAGRA:2022twx}. Future
space-based interferometers such as LISA~\cite{LISA:2017pwj}, TianQin~\cite{TianQin:2015yph} and Taiji~\cite{Hu:2017mde} will probe the millihertz
window. Pulsar timing arrays (PTAs) are sensitive to nanohertz
frequencies~\cite{NANOGrav:2023gor,Janssen:2014dka}. Recently, several PTA
collaborations --- NANOGrav~\cite{NANOGrav:2023gor}, EPTA/InPTA~\cite{EPTA:2023sfo,InPTA:2023gpu},
PPTA~\cite{Reardon:2023gzh} and CPTA~\cite{Xu:2023wog} --- have reported evidence for a stochastic GW background. While the most conservative
interpretation is a background from supermassive black-hole
binaries~\cite{NANOGrav:2023hde,Sato-Polito:2023gym,Ellis:2023dgf}, a cosmological origin cannot be excluded, and
identifying it requires precision predictions for every candidate source~\cite{NANOGrav:2023hvm,Ellis:2023oxs}.

One class of cosmological sources has attracted particular attention: scalar-induced
gravitational waves (SIGWs) (see Refs.~\cite{Domenech:SIGWRe, Yuan:2021qgz} for a review) generated at second order in perturbation theory by the same
primordial curvature perturbations that seeded the anisotropy in the cosmic microwave background (CMB) and the large scale structure of our Universe~\cite{Matarrese:1992rp,Matarrese:1993zf,Matarrese:1997ay,Ananda:2006af,Baumann:2007zm,Saito:2008jc}. While the CMB, baryon acoustic oscillations (BAO) and the Lyman-$\alpha$ forest constrain the curvature power spectrum only at large scales~\cite{Planck:2018vyg}, SIGWs can probe it at small scales. Explaining the PTA data with SIGWs requires the small-scale power to be enhanced well above
the CMB normalization. This links SIGW searches to primordial black hole formation~\cite{Franciolini:2023pbf,Ellis:2023oxs,Wang:2024nmd} and to a rich variety of enhancement mechanisms in the early Universe and in inflation~\cite{Kinney:2005vj,Inoue:2001zt,Cai:2021zsp}.
These features have made SIGWs one of the most popular explanations of the PTA
signal~\cite{Franciolini:2023wjm}.

The standard SIGW calculation assumes radiation domination with a constant equation of state $w$ and sound speed $c_s$ ($w=c_s^2=1/3$). However, this assumption is not realized in the real Universe. As the temperature of the Universe drops, the relativistic degrees of freedom in the thermal bath, as well as $w$ and $c_s$, change when the temperature passes a certain energy scale that triggers the corresponding dynamics (e.g., phase transitions). Within the Standard Model (SM), there are processes such as electroweak (EW) crossover, QCD crossover, and electron-positron annihilation. 
During each process, the plasma is a mixture of
relativistic and non-relativistic species. Therefore, both \(w\) and \(c_s^2\) depart from \(1/3\) with non-trivial time/temperature dependence. In models beyond the SM (BSM), there can be more options to change $w$ and $c_s$: modifications to existing dynamics (such as EW baryogenesis where the EW phase transition becomes first-order~\cite{KUZMIN198536,Shaposhnikov:1986jp,Shaposhnikov:1987tw,Cohen:1993nk,Trodden:1998ym,Riotto:1999yt,Dine:2003ax,Morrissey:2012db}); additional phase transitions (such as in Composite Higgs models~\cite{Creminelli:2001th, Randall:2006py, Nardini:2007me, Konstandin:2010cd, Konstandin:2011dr, Bunk:2017fic, Baratella:2018pxi,Bruggisser:2018mrt, Megias:2018sxv, Agashe:2019lhy,Fujikura:2019oyi, Agashe:2020lfz, Ares:2020lbt,Agrawal:2021alq, Levi:2022bzt, Csaki:2023pwy,Eroncel:2023uqf,Mishra:2023kiu,Mishra:2024ehr} or in dark sectors~\cite{Bai:2013xga,Kribs:2016cew,Cline:2021itd,Shelton:2010ta,DUTTA2011364,Holthausen:2013ota,Schwaller:2015tja,Hall:2019rld,Bai:2021ibt,Bottaro:2021aal,Pasechnik:2023hwv}); departure from radiation domination (such as early matter domination~\cite{Assadullahi:2009nf,Lidsey:2001nj,Hidalgo:2011fj,Inomata:2025wiv,Johnson:2008se,Alabidi:2012ex,Alabidi:2013lya,Inomata:2019ivs,Inomata:2020lmk,Inomata:2019zqy,Pearce:2023kxp,Kumar:2024hsi,Zhou:2026iry,Papanikolaou:2024kjb,He:2026neq,Gouttenoire:2026wcd,Zeng:2025tno,Papanikolaou:2020qtd,Domenech:2020ssp,Mansour:2026sdx}). The thermal history with time-dependent background will impact SIGWs in a non-trivial way. Therefore, the SIGW spectrum encodes the information of dynamics (both in the SM and BSM models) in the early Universe, some of which are hard to calculate theoretically. For example, the QCD crossover in the SM (and BSM models like Composite Higgs models and confining dark sectors) involves non-perturbative dynamics that generally requires lattice calculations. Measurements of $w$ and $c_s$ associated with these phase transitions will be critical to test the SM and BSM models. Interestingly, the SIGW signals associated with QCD crossover and TeV scale phase transitions naturally fall in the detectable window of PTAs and the planned space-based experiments (such as LISA, Taiji and TianQin), respectively. Therefore, a precise calculation of SIGW in a general time-dependent thermal bath is important for testing the SM and BSM models, as well as building a robust connection with data from existing and upcoming GW observations.

The SIGW spectrum associated with QCD crossover in the SM has been studied in Refs.~\cite{Hajkarim:2019nbx,Abe:2020sqb,Abe:2023yrw,Franciolini:2026ner,Bhaumik:2025vlb,Pritchard:2024vix} with the input of $w$ and $c_s$ from lattice QCD calculation~\cite{Borsanyi:2016} {(see also Refs.~\cite{Watanabe:2006qe,Saikawa:2018rcs,Feng:2026xwp,Gonin:2026tyu,Pritchard:2025pcn,Escriva:2024ivo,Escriva:2023nzn,Dalianis:2020cla,Liu:2023hpw,Zhu:2023gmx,Domenech:2024rks,Escriva:2022bwe,Musco:2023dak,Liu:2023pau,Balaji:2023ehk,Domenech:2020kqm,Domenech:2019quo,Byrnes:2018clq,Carr:2019kxo,Ferreira:2026uzi} for studies of the thermal history)}. However, due to uncertainties in the lattice calculation and fitting functions, the resultant $w$ and $c_s$ do not conserve entropy~\cite{Abe:2020sqb}. The standard SIGW calculation relies on the assumption of an adiabatic bath with conservation of entropy. Therefore, there is a potential concern whether special features of the resultant SIGW spectrum are an artifact of entropy non-conservation. Moreover, the standard way of setting initial conditions for metric perturbations is to use initial conditions in standard radiation domination ($w=c_s^2=1/3$) at some initial time. However, for the calculation of certain modes, the initial time might fall in periods where $w$ and $c_s$ have significant changes. As we will show in Sec.~\ref{subsec:initial condition}, the standard radiation initial conditions might lead to an artificial enhancement in the metric perturbation and induced GW spectrum. To address the above issues, we first revisit SIGWs in a general time-dependent background. We introduce a toy model with conservation of entropy, whose $w$ and $c_s$ have similar shapes to those of phase transitions. Moreover, we derive general initial conditions for metric perturbations obtained from solutions to the superhorizon evolution in a time-dependent background. We calculate the induced GW spectra based on the toy model, which exhibit qualitatively the same features as the previous results of QCD crossover, such as the step-like shape due to the entropy dilution factor, phase delay, a peak and a dip around the transition. This justifies the robustness of those features associated with phase transitions. We then apply this formalism to the SM case involving QCD crossover, EW crossover and electron-positron annihilation. To address the entropy non-conservation issue of previous \(w\) and \(c_s\) during QCD crossover, we construct new values of \(w\) and \(c_s\) within error bars of lattice results that explicitly conserve entropy. This entropy-conserving QCD input therefore provides a well-defined setting for computing the corresponding SIGWs. The GW spectrum from the full SM thermal history exhibits a combination of features discussed above from each transition. Interestingly, the most prominent feature of the SIGW spectrum from the SM is the step-like shape around $10$ nHz associated with QCD crossover, which lies in the observational window of PTAs. We then test SIGW from the SM with the 15-year NANOGrav data with two primordial curvature power spectra: a delta-function spectrum and a broken power law. We find, for both power spectra, that current data have only a very mild preference over the standard radiation case, and the constraints also change slightly. However, the best fit of the peak frequency of the delta-function power spectrum shifts significantly, which can be distinguished by future GW observations in the $\mu\textrm{Hz}$ window.
We further study an example of phase transitions beyond the SM (in particular, composite Higgs models) based on our toy model. We find that such phase transitions around TeV can be detected by future space-based GW experiments (such as LISA, Taiji and TianQin). Therefore, GW spectrum observations across many orders of magnitude in frequency serve as a powerful probe of phase transitions in the SM and BSM models in the early Universe.

The rest of this paper is organized as follows. In Sec.~\ref{sec:formalism}, we review
SIGWs on a general adiabatic single-fluid background, discuss different choices of the primordial curvature power spectra, and derive exact initial conditions for the metric perturbation based on its superhorizon evolution. In Sec.~\ref{sec:eos_PT}, we show the equation of state and sound speed for a toy model, the SM background, and BSM models.
In Sec.~\ref{sec:thermal}, we first compute the induced GW spectra based on the toy
model, and then study the full Standard Model thermal history. We also show examples of phase transitions from BSM models. In Sec.~\ref{sec:pta}, we analyze the SIGW spectra from the full SM thermal history based on the existing NANOGrav 15-year data, and discuss the potential of probing BSM phase transitions with future GW observations. Finally, we conclude in
Sec.~\ref{sec:concl}.

\section{Scalar Induced Gravitational Waves in a Time-dependent Background}
\label{sec:formalism}

\subsection{General Formalism}

During a cosmological phase transition (including a crossover), the effective number of relativistic
degrees of freedom $g_*$ of the thermal bath changes as the temperature drops. We approximate the thermal bath as a single fluid in local thermal equilibrium with negligible chemical potentials. Its background energy density $\rho$ and entropy density $s$ depend on a common
temperature, and thus can be parameterized by $g_*$ and the effective entropy
degrees of freedom $g_{*s}$, respectively:
\begin{equation}
    \rho(T)=\frac{\pi^2}{30}g_*(T)T^4,
    \qquad
    s(T)=\frac{2\pi^2}{45}g_{*s}(T)T^3,
    \label{eq:thermo}
\end{equation}
with \(Ts=\rho+p\) where $p$ denotes the background pressure density. The equation-of-state parameter \(w=p/\rho\) and the adiabatic sound speed squared
\(c_s^2=(\d p/\d T)/(\d\rho/\d T)\) can then be expressed as
\begin{equation}
    w(T)=\frac{4g_{*s}(T)}{3g_*(T)}-1,
    \qquad
    c_s^2(T)=
    \frac{4\left[\d g_{*s}(T)/\d\ln T +4g_{*s}(T)\right]}
    {3\left[\d g_{*}(T)/\d\ln T+4g_*(T)\right]}-1 .
    \label{eq:wcs_thermo}
\end{equation}
If both effective degrees of freedom are constant and equal to each other, we get \(c_s^2=w=1/3\), corresponding to standard radiation domination.

The homogeneous background is determined by
\begin{equation}
    \rho'=-3\mathcal{H}(1+w)\rho,
    \qquad
    \mathcal{H}=\frac{a'}{a},
    \qquad
    \mathcal{H}^2=\frac{8\pi G}{3}a^2\rho,
    \label{eq:background}
\end{equation}
where $\mathcal H$ denotes the conformal Hubble parameter, $a$ is the scale factor, and the prime denotes a derivative with respect to the conformal time $\eta$. Using \(c_s^2=\d p/\d\rho\), the continuity equation gives the equivalent temperature equation
\begin{equation}
    \frac{T'}{T}=-3\mathcal{H}c_s^2[T(\eta)].
    \label{eq:T_eta}
\end{equation}
For a thermodynamically consistent adiabatic equation of state, Eq.~\eqref{eq:T_eta} is equivalent to
\(s[T(\eta)]a^3(\eta)=\mathrm{constant}\). Thus there are two equivalent parameterizations of the same background: solving \(\rho(\eta)\) together with Eq.~\eqref{eq:background}, and solving \(T(\eta)\) together with Eq.~\eqref{eq:T_eta} and the Friedmann equation. To fix the normalization, we set the present-day temperature $T_0=T_{\gamma,0}$ and present-day scale factor $a_0=1$.
The equation of state \(w\) and sound speed \(c_s^2\) then become functions of conformal time through \(T(\eta)\). The
single-fluid description is particularly convenient here: the whole transition reduces to
the two functions \(w(\eta)\) and \(c_s^2(\eta)\), without decomposing the plasma into
individual particle species.
A time-dependent \(w(\eta)\) therefore modifies the expansion history, so that both \(a(\eta)\) and \(\mathcal{H}(\eta)\) deviate from \(a\propto\eta\) and \(\mathcal{H}\propto1/\eta\). A mistaken use of the radiation-domination relation \(\eta\propto1/T\) would displace the transition in conformal time and lead to a non-physical reduction of the entropy change.

On such a background, scalar perturbations source tensor modes at second order in
perturbation theory. We work in conformal time \(\eta\), quoted in units of Mpc, and in the conformal Newtonian gauge~\cite{Domenech:2025ccu, Domenech:2020xin}. We neglect anisotropic stress at first and second order~\cite{Kodama:1986fg,Kodama:1986ud,Ma:1995ey}:
\begin{equation}
    ds^2=a^2(\eta)\left[-(1+2\Phi)d\eta^2+
    \left((1-2\Phi)\delta_{ij}+\frac12 h_{ij}\right)dx^idx^j\right],
    \label{eq:metric}
\end{equation}
where \(h_{ij}\) is the transverse-traceless gravitational-wave perturbation and
\(\Phi\) is the scalar metric perturbation. 
Since in this work we only focus on adiabatic/curvature perturbations, the metric perturbation $\Phi$ can be written in the Fourier space as \(\Phi_{\mathbf{k}}(\eta)=\Phi_k(\eta)\,\zeta_{\mathbf{k}}\), where \(\zeta_{\bf k}\) is the primordial curvature perturbation and $\Phi_k(\eta)$ is the transfer function, which obeys the following evolution equation:
\begin{equation}
    \Phi_k''+3\mathcal{H}(1+c_s^2)\Phi_k'
    +\left[c_s^2k^2+3\mathcal{H}^2(c_s^2-w)\right]\Phi_k=0.
    \label{eq:scalar_eom}
\end{equation}
The tensor mode sourced at second order by scalar perturbations satisfies
\begin{equation}
    \left[\frac{\partial^2}{\partial\eta^2}+k^2-
    \frac{1-3w(\eta)}{2}\mathcal{H}^2\right]\left[a(\eta)h_{\mathbf{k}}(\eta)\right]
    =4a(\eta)S_{\mathbf{k}}(\eta).
    \label{eq:tensor_eom}
\end{equation}
The scalar source is
\begin{align}
    S_{\mathbf{k}}(\eta)
    =&\int\frac{d^3\tilde{\vec k}}{(2\pi)^3}
    e^{ij}(\mathbf{k})\tilde{k}_i\tilde{k}_j
    \bigg[2\Phi_{\tilde{\mathbf{k}}}\Phi_{\mathbf{k}-\tilde{\mathbf{k}}}
    \nonumber\\
    &\quad+\frac{4}{3[1+w(\eta)]}
    \left(\Phi_{\tilde{\mathbf{k}}}+\frac{\Phi'_{\tilde{\mathbf{k}}}}{\mathcal{H}}\right)
    \left(\Phi_{\mathbf{k}-\tilde{\mathbf{k}}}+\frac{\Phi'_{\mathbf{k}-\tilde{\mathbf{k}}}}{\mathcal{H}}\right)\bigg].
    \label{eq:source}
\end{align}
Here \(e^{ij}(\mathbf{k})\) is the transverse-traceless polarization tensor projecting onto
the two tensor polarizations. The equation of state $w$ appears explicitly in tensor
propagation and in the source, whereas the sound speed $c_s^2$ enters through the scalar transfer functions.

To construct the retarded Green's function, we first solve the homogeneous
equation associated with the rescaled tensor mode \(g_k=a h_k\)~\cite{Kohri:2018awv},
\begin{equation}
    \left[\frac{\partial^2}{\partial\eta^2}+k^2-
    \frac{1-3w(\eta)}{2}\mathcal{H}^2\right]g_{k,i}(\eta)=0,
    \qquad i=1,2.
    \label{eq:tensor_homogeneous}
\end{equation}
Let \(g_{k,1}\) and \(g_{k,2}\) be two independent solutions, with the
constant Wronskian \(W_k=g_{k,1}g_{k,2}'-g_{k,1}'g_{k,2}\). Their combination gives
\begin{equation}
    G_k(\eta,\tilde\eta)=\Theta(\eta-\tilde\eta)
    \frac{g_{k,1}(\tilde\eta)g_{k,2}(\eta)
    -g_{k,2}(\tilde\eta)g_{k,1}(\eta)}{W_k},
    \label{eq:green}
\end{equation}
so that \(G_k(\tilde\eta,\tilde\eta)=0\) and
\(\partial_\eta G_k(\eta,\tilde\eta)|_{\eta=\tilde\eta^+}=1\).
The sourced tensor solution is
\begin{equation}
    h_{\mathbf{k}}(\eta)=\frac{4}{a(\eta)}\int_0^\eta d\tilde{\eta}\,
    G_k(\eta,\tilde{\eta})a(\tilde{\eta})S_{\mathbf{k}}(\tilde{\eta}).
    \label{eq:tensor_solution}
\end{equation}
For Gaussian perturbations, and with the convention of Eq.~\eqref{eq:power_convention},
the induced tensor spectrum becomes
\begin{align}
    \mathcal{P}_h(k,\eta)
    =&16
    \int_{|k_1-k_2|\leq k\leq k_1+k_2}d\ln k_1\,d\ln k_2\,
    I^2(k,k_1,k_2,\eta)
    \nonumber\\
    &\times\frac{\left[k_1^2-\frac{(k^2-k_2^2+k_1^2)^2}{4k^2}\right]^2}
    {k_1k_2k^2}\mathcal{P}_\zeta(k_1)\mathcal{P}_\zeta(k_2),
    \label{eq:Ph_general}
\end{align}
where \(k_1=\tilde{k}\) and \(k_2=|\mathbf{k}-\tilde{\mathbf{k}}|\) are the two momenta entering the convolution. Here
\begin{align}
    I(k,k_1,k_2,\eta)=\frac{k^2}{a(\eta)}\int_0^\eta d\tilde{\eta}\,a(\tilde{\eta})G_k(\eta,\tilde{\eta})
    \bigg[\Phi_{k_1}(\tilde{\eta})\Phi_{k_2}(\tilde{\eta})
    \nonumber\\
    +\frac{2}{3[1+w(\tilde{\eta})]}
    \left(\Phi_{k_1}+\frac{\Phi'_{k_1}}{\mathcal{H}}\right)
    \left(\Phi_{k_2}+\frac{\Phi'_{k_2}}{\mathcal{H}}\right)\bigg].
    \label{eq:kernel_general}
\end{align}
The dimensionless power spectrum is defined as
\begin{equation}
    \langle X_{\mathbf{k}}X_{\mathbf{k}'}\rangle=(2\pi)^3\delta^{(3)}(\mathbf{k}+\mathbf{k}')
    \frac{2\pi^2}{k^3}\mathcal{P}_X(k),\qquad X=h,\zeta .
    \label{eq:power_convention}
\end{equation}

For comparison, in standard radiation domination, \(w=c_s^2=1/3\) and \(a\propto\eta\).
The regular solution of Eq.~\eqref{eq:scalar_eom}, normalized to
\(\Phi_k\to 2/3\) as \(k\eta\to 0\), is
\(\Phi_k(\eta)=2j_1(y)/y\), where \(y\equiv k\eta/\sqrt{3}\) and \(j_1\)
is the spherical Bessel function of the first kind.
The two independent solutions of Eq.~\eqref{eq:tensor_homogeneous} can then be
chosen as \(\propto\cos(k\eta)\) and \(\propto\sin(k\eta)\). Equation~\eqref{eq:green}
gives the retarded Green's function
\(G_k(\eta,\tilde\eta)=\Theta(\eta-\tilde\eta)
\sin[k(\eta-\tilde\eta)]/k\).
With these solutions, Eq.~\eqref{eq:kernel_general} reduces to the standard radiation SIGW kernel~\cite{Domenech:SIGWRe}.
The changes induced by a time-dependent background originate from three parts:
\begin{enumerate}
\item \(c_s^2(\eta)\), \(w(\eta)\) and \({\cal H}(\eta)\) change the equation of the scalar transfer function in Eq.~\eqref{eq:scalar_eom};
\item \(w(\eta)\) and \({\cal H}(\eta)\) change the effective potential in the tensor Green's function in Eq.~\eqref{eq:green};
\item \(w(\eta)\) and \({\cal H}(\eta)\) change the coefficients in the time integration of the kernel function in Eq.~\eqref{eq:kernel_general}.
\end{enumerate}
After the source has settled, we evaluate the GW energy-density fraction at a time \(\eta_*\) well inside the horizon, when the tensor modes propagate as radiation:
\begin{equation}
    \Omega_{\rm GW}(k,\eta_*)\equiv\frac{1}{\rho}\frac{\d \rho_{\rm GW}}{\d \ln k}=\frac{1}{24}
    \left(\frac{k}{\mathcal{H}(\eta_*)}\right)^2\overline{\mathcal{P}_h(k,\eta_*)} ,
    \label{eq:omega_generation}
\end{equation}
where the overline indicates the time average around $\eta_*$.
To obtain the GW spectrum today, we need to consider the dilution due to heating of the photon bath from the change of $g_{*s}$:
\begin{equation}
    \Omega_{\textrm{GW},0}(k)h^2=
    \Omega_{r,0}h^2\frac{g_{*}(\eta_*)}{g_{*,0}}
    \left(\frac{g_{*s,0}}{g_{*s}(\eta_*)}\right)^{4/3}\Omega_{\rm GW}(k,\eta_*),
    \label{eq:redshift_general}
\end{equation}
where \(\Omega_{r,0}\) is the present-day fraction of radiation density and \(h\equiv H_0/(100\, \textrm{km/s/Mpc})\) is the dimensionless Hubble constant.
Considering \(g_{*}(\eta_*)\simeq g_{*s}(\eta_*)\), and taking the present-day values
\(g_{*,0}=3.38\), \(g_{*s,0}=3.94\) and \(\Omega_{r,0}h^2=4.18\times10^{-5}\)~\cite{Planck:2018vyg}, one can obtain
\begin{equation}
    \Omega_{\textrm{GW},0}(k)h^2\simeq1.67\times10^{-5}
    \left(\frac{g_{*s}(\eta_*)}{100}\right)^{-1/3}\Omega_{\rm GW}(k,\eta_*).
    \label{eq:redshift_approx}
\end{equation}
The precise calculation of the induced GW spectrum is computationally intensive due to the lack of analytical solutions in a general time-dependent background. One needs to solve the scalar perturbation and tensor Green's function for each $k$ mode. The GW spectrum further requires integrals over wide ranges in $k$ space and in the time domain. However, since data analysis focuses on fitting or setting constraints on parameters in the primordial power spectrum $\mathcal P_\zeta(k)$, we only need to evaluate the source kernel $I(k,k_1,k_2,\eta)$, which is the most computationally challenging, once for a given thermal history. In this work, we discretize the $k$ space to calculate $\Phi_k$ and $G_k$ for each $k$ mode. We then tabulate the time-averaged source kernel ${\cal K}(k,k_1,k_2)\equiv(k/{\mathcal H(\eta_*)})^2\ \overline{I^2(k,k_1,k_2,\eta_*)}$ for different values of $k,k_1,k_2$. The induced GW spectrum for different $\mathcal P_\zeta$ can be simply obtained with the documented ${\cal K}$.
The details of our numerical computation program can be found in Appendix~\ref{app:numerical}.

\paragraph{Primordial Power Spectrum.}
The spatial distribution of the primordial curvature perturbation is encoded in the curvature power spectrum \(\mathcal P_\zeta\), which depends on the underlying model. In this study, we use three exemplary spectra: a flat spectrum, a delta-function spectrum, and a broken power law (BPL) spectrum. 

The flat spectrum is
\begin{equation}
    \mathcal{P}_{\zeta}(k)=A,
    \label{eq:P_flat}
\end{equation}
where $A$ denotes the amplitude. We note that $A$ is treated as a free parameter for later GW analysis, not an extrapolation of the large-scale amplitude measured from the cosmic microwave background. A flat spectrum across a wide range in $k$ can be achieved in certain curvaton models (see, e.g., Ref~\cite{Kumar:2024hsi}). Moreover, this flat spectrum
is useful to show imprints of the time-dependent background at different scales.

The delta-function spectrum has the form
\begin{equation}
    \mathcal{P}_{\zeta}(k)=A\,\delta(\ln k-\ln k_0),
    \label{eq:P_delta_draft}
\end{equation}
where $A$ is the amplitude and $k_0$ is the characteristic wavenumber. While no physical model provides an exactly delta-function power spectrum, it approximates a narrow peak in the power spectrum and allows us to observe the effect of the time-dependent background at each characteristic wavenumber. Moreover, the delta-function power spectrum is widely used to obtain general constraints on any power spectra assuming no correlations between different $k$ modes.

The BPL spectrum has the form
\begin{equation}
    \mathcal{P}_{\zeta}(k)=A\begin{cases}
    (k/k_0)^3,& k\leq k_0,\\
    1,& k>k_0,
    \end{cases}
    \label{eq:P_BPL_draft}
\end{equation}
where $k_0$ denotes the scale where the BPL transition occurs. The $k^3$ dependence for small $k$ is generic for a source with finite-range spatial correlations on scales larger than its correlation length (the white-noise limit, see e.g.,~\cite{Graham:2015rva,Irsic:2019iff,Feix:2020txt,Amin:2022nlh,Redi:2022llj,Elor:2023xbz,Graham:2024hah,deKruijf:2024voc,Ling:2024qfv,Amin:2025dtd,Cai:2019cdl}). If the observational data are mostly from this $k^3$ tail (i.e., for large $k_0$), the thermal history correction can be applied to any model that has a $k^3$ spectrum up to rescaling. One concrete realization of this BPL spectrum is a non-thermal phase transition during inflation~\cite{Buckley:2024nen,Barir:2022kzo}.

\subsection{Initial Conditions for the Metric Perturbation}
\label{subsec:initial condition}

To obtain the general form of initial conditions for the metric perturbation $\Phi_{\vec k}(\eta)$, we need to study its superhorizon evolution.
For the adiabatic mode, the conserved quantity outside the horizon is the curvature perturbation \(\zeta_{\bf k}\),
not the metric perturbation. In the absence of anisotropic stress, these two are related as~\cite{Gordon:2000hv}
\begin{equation}
    \zeta_{\bf k}=\Phi_{\bf k}
    +\frac{2}{3[1+w(\eta)]}
    \left(\frac{\Phi_{\bf k}'}{\mathcal H}+\Phi_{\bf k}\right),
    \label{eq:zeta_Phi_superhorizon}
\end{equation}
up to gradient corrections of order \(k^2/\mathcal H^2\). For a constant equation of
state, the regular mode has \(\Phi_{\bf k}'=0\), and Eq.~\eqref{eq:zeta_Phi_superhorizon}
reduces to the familiar relation
\begin{equation}
    \Phi_{\bf k}=\frac{3(1+w)}{5+3w}\,\zeta_{\bf k}~~~~~(\textrm{for a constant}\, w).
    \label{eq:Phi_zeta_constant_w}
\end{equation}
For a general time-dependent $w(\eta)$, Eq.~\eqref{eq:zeta_Phi_superhorizon} becomes the
first-order evolution equation
\begin{equation}
    \Phi_{\bf k}'
    +\frac{5+3w}{2}\mathcal H\Phi_{\bf k}
    =\frac{3}{2}(1+w)\mathcal H\zeta_{\bf k}.
    \label{eq:Phi_zeta_evolution}
\end{equation}
The nonvanishing derivative makes \(\Phi_{\bf k}\) sensitive to the preceding expansion
history rather than to the instantaneous value of \(w(\eta)\) alone.

For the adiabatic single-fluid backgrounds considered here,
\begin{equation}
    c_s^2=\frac{p'}{\rho'}
    =w-\frac{w'}{3\mathcal H(1+w)},
    \label{eq:adiabatic_cs_w}
\end{equation}
and the nonadiabatic pressure perturbation vanishes. We note that \(c_s^2\ne w\) during a transition does not break the conservation of \(\zeta_{\bf k}\). Conservation of
\(\zeta_{\bf k}\) fixes the normalization of the regular mode, while
Eq.~\eqref{eq:Phi_zeta_evolution} determines how \(\Phi_{\bf k}\) evolves in that background.

\begin{figure}[h]
    \centering
    \includegraphics[width=0.7\linewidth]{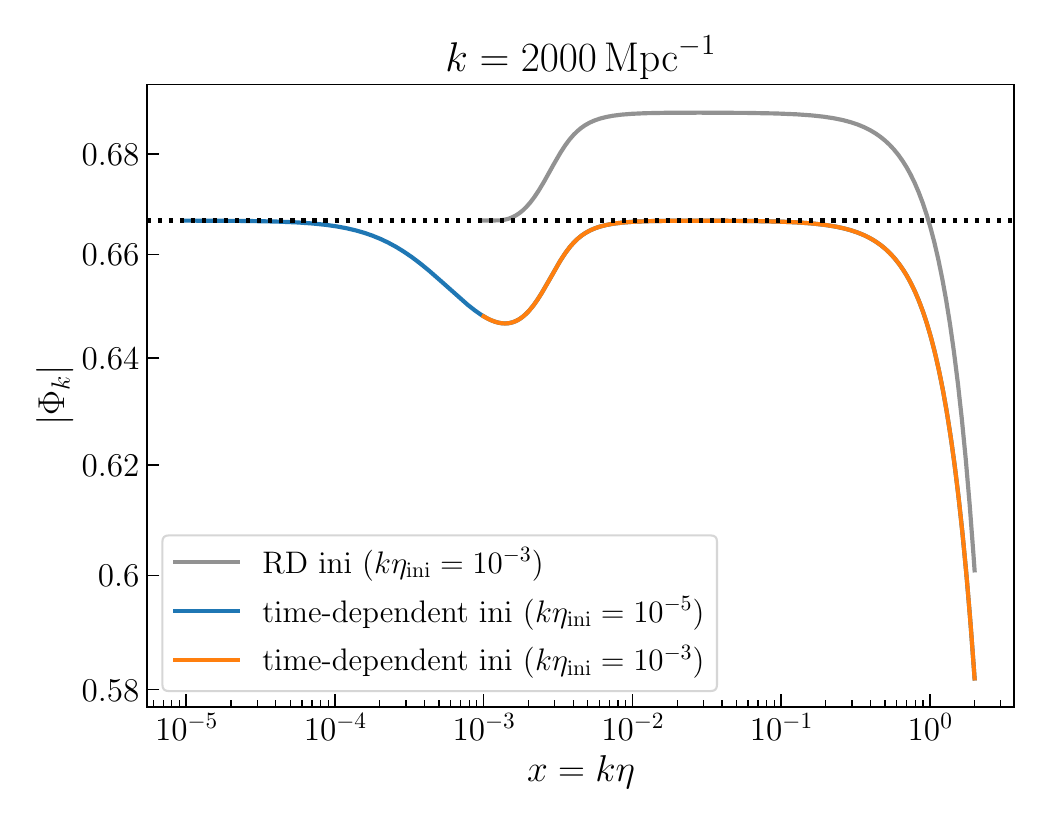}
    \caption{Superhorizon evolution of the transfer function $\Phi_k$ for different choices of initial conditions for
    \(k=2000\,\mathrm{Mpc}^{-1}\) mode based on the toy model discussed in Sec.~\ref{sec:toy_model} with $r_g=10$ and $m=800\,\textrm{MeV}$. The blue and orange curves use the time-dependent initial conditions (Eq.~\ref{eq:Phi_super_conformal_draft}) at \(k\eta_{\rm ini}=10^{-5}\) and
    \(10^{-3}\), respectively. The gray curve instead uses the standard radiation initial condition at \(k\eta_{\rm ini}=10^{-3}\); the black dotted line marks \(2/3\).}
    \label{fig:initial condition}
\end{figure}

To obtain a general solution of $\Phi_{\vec k}(\eta)$ in the superhorizon limit, we first assume that the thermal bath is in standard radiation domination (\(w=c_s^2=1/3\)) at $\eta\to0$.
Equation~\eqref{eq:Phi_zeta_constant_w} then gives
\(\Phi_{\bf k}(\eta\to0)=2\zeta_{\bf k}/3\). This condition selects the regular adiabatic mode and
excludes the independent decaying solution. We subsequently propagate this mode through
the time-dependent background while it remains outside the horizon.
As derived in Appendix~\ref{app:superhorizon}, the resulting scalar transfer function,
defined by \(\Phi_{\bf k}(\eta)=\Phi_k(\eta)\zeta_{\bf k}\), and its derivative are
\begin{equation}
    \Phi_k(\eta)=\left(1-\frac{\mathcal{H}}{a^2}
    \int_0^\eta a^2(\tilde\eta)\,\mathrm{d}\tilde\eta\right)~ ,~\Phi'_k(\eta)=-\mathcal H-(\mathcal H'-2\mathcal H^2)\left(\frac{1}{a^2}
    \int_0^\eta a^2(\tilde\eta)\,\mathrm{d}\tilde\eta\right).
    \label{eq:Phi_super_conformal_draft}
\end{equation}
This expression assumes adiabatic perturbations in a background with sound speed
satisfying Eq.~\eqref{eq:adiabatic_cs_w}, neglecting anisotropic stress and gradient terms. 
We use both its value $\Phi_k(\eta_{\rm ini})$ and
its derivative $\Phi_k'(\eta_{\rm ini})$ as initial conditions for the transfer function at \(k\eta_{\rm ini}=10^{-3}\), for which
\(k/\mathcal H(\eta_{\rm ini})\ll1\). In the standard radiation domination ($\mathcal H= \eta^{-1}\propto a^{-1}$), Eq.~\eqref{eq:Phi_super_conformal_draft} reduces to the standard radiation initial condition $\Phi_k(\eta_{\rm ini})=2/3$ and $\Phi'_k(\eta_{\rm ini})=0$. However, using \(\Phi_k(\eta_{\rm ini})=2/3\) in a time-dependent background would lead to an artificial enhancement for certain $k$ modes. To view this, we apply it to the case with phase transitions in the SM in Fig.~\ref{fig:initial condition}. As will be seen later, $w(\eta)$ and $c_s^2(\eta)$ exhibit a dip feature during phase transitions, and return to $w(\eta)=c_s^2(\eta)=1/3$ before and after each transition. As a result, $\Phi_k$ develops a dip feature during phase transitions outside the horizon according to Eq.~\eqref{eq:Phi_super_conformal_draft}. Since we fix $k\eta_{\rm ini}=10^{-3}$, the initial time varies for different $k$, some of which might fall in the period where $w$ and $c_s^2$ have significant changes. We show an example of the superhorizon evolution of $\Phi$ with \(k=2\times 10^3\,\mathrm{Mpc}^{-1}\) in Fig.~\ref{fig:initial condition} based on the toy model of $w$ and $c_s$ discussed in Sec.~\ref{sec:toy_model}. Given the parameters of the toy model ($r_g=10$ and $m=800\,\textrm{MeV}$), the initial time of this mode \(\eta_{\rm ini}=10^{-3}/k=5\times 10^{-7}\,\mathrm{Mpc}\) is around the middle of the transition. Using the standard radiation initial condition $\Phi_k(\eta_{\rm ini})=2/3$ at $\eta_{\rm ini}$ will lead to an overestimate of the amplitude at a later time due to subsequent enhancement when $\omega$ and $c_s^2$ increase from $<1/3$ to $1/3$. Considering SM phase transitions, this artificial enhancement can be as large as $10\%$ for QCD crossover, which will also lead to an artificial bump around \(k\sim10^3\,\mathrm{Mpc}^{-1}\) in the induced GW spectrum (see Appendix~\ref{app:superhorizon}).

\section{Equation of State and Sound Speed from Phase Transitions}
\label{sec:eos_PT}

The thermal history of the early Universe provides concrete realizations of a time-dependent background. The formalism of SIGW presented in the previous section needs the input of $w$ and $c_s^2$ from different models of the thermal history. We first introduce a toy model for a single phase transition with the conservation of entropy. We then move on to study phase transitions in the SM, with a new form of $w$ and $c_s^2$ based on a lattice QCD calculation, which conserves entropy. Finally, we analyze an example of phase transitions in BSM models: the confinement phase transition in composite Higgs models.

\subsection{Toy Model for Phase Transitions}
\label{sec:toy_model}
In this section, we introduce a toy model for a single phase transition with conservation of entropy that can be directly applied to calculate SIGWs. The reason for studying this toy model is the following. The $w$ and $c_s^2$ needed for SIGW calculations are difficult to calculate for general phase transitions. Moreover, for the SM ones obtained from lattice calculation, there are multiple transitions that tend to mix effects from different transitions. Therefore, it is useful to have a simple toy model that can isolate the impact of a single phase transition and be applicable to general phase transitions.

The toy model describes a coupled fluid consisting of massive and massless particles, with an explicit realization given in Ref.~\cite{Aloni:2022} and summarized in Appendix~\ref{app:stepped}.
When the temperature drops below the mass threshold, massive particles dump their entropy into massless species, with the total entropy conserved. This process results in an increase in temperature and total energy density, the same effect as the heating of the photon bath due to electron-positron annihilation. During the transition, the fluid is a mixture of massive and massless particles. As a result, both \(w\) and \(c_s^2\) drop below their radiation values ($1/3$), and return to \(1/3\) once the massive states have annihilated. We note that this is not a model for real phase transitions or crossovers. However, since there are also some degrees of freedom that become massive during phase transitions and later annihilate into lighter species, the toy model shares similar features, such as dips in $w$ and $c_s^2$, with those in phase transitions. 

In the toy model, we consider massive particles with the same mass $m$. We define \(x=m/T\) as the ratio of the particle mass to the fluid temperature $T$, and \(r_g=(g_{*,\rm UV}-g_{*,\rm IR})/g_{*,\rm IR}\) as the ratio of the massive degrees of freedom to those remaining relativistic. With these variables the thermodynamics gives
\begin{equation}
    w(x)=\frac{1+r_g\hat p(x)}{3\left[1+r_g\hat\rho(x)\right]},
    \label{eq:w_stepped}
\end{equation}
where \(\hat\rho(x)\) and \(\hat p(x)\) are dimensionless energy and pressure densities:
\begin{equation}
    \hat\rho(x)=\frac{1}{6}\int_0^\infty \mathrm{d}u\,u^2\sqrt{u^2+x^2}\,
    e^{-\sqrt{u^2+x^2}},
    \qquad
    \hat p(x)=\frac{1}{6}\int_0^\infty \mathrm{d}u\,\frac{u^4}{\sqrt{u^2+x^2}}\,
    e^{-\sqrt{u^2+x^2}}.
    \label{eq:rhoprime_integrals}
\end{equation}
The sound speed follows from
\(c_s^2=\d p/\d\rho\) in an analogous closed form. The full construction, together with
the entropy-conservation relation that fixes \(x\) as a function of the scale factor, is
collected in Appendix~\ref{app:stepped}.

\begin{figure}[ht]
    \centering
    \includegraphics[width=0.65\linewidth]{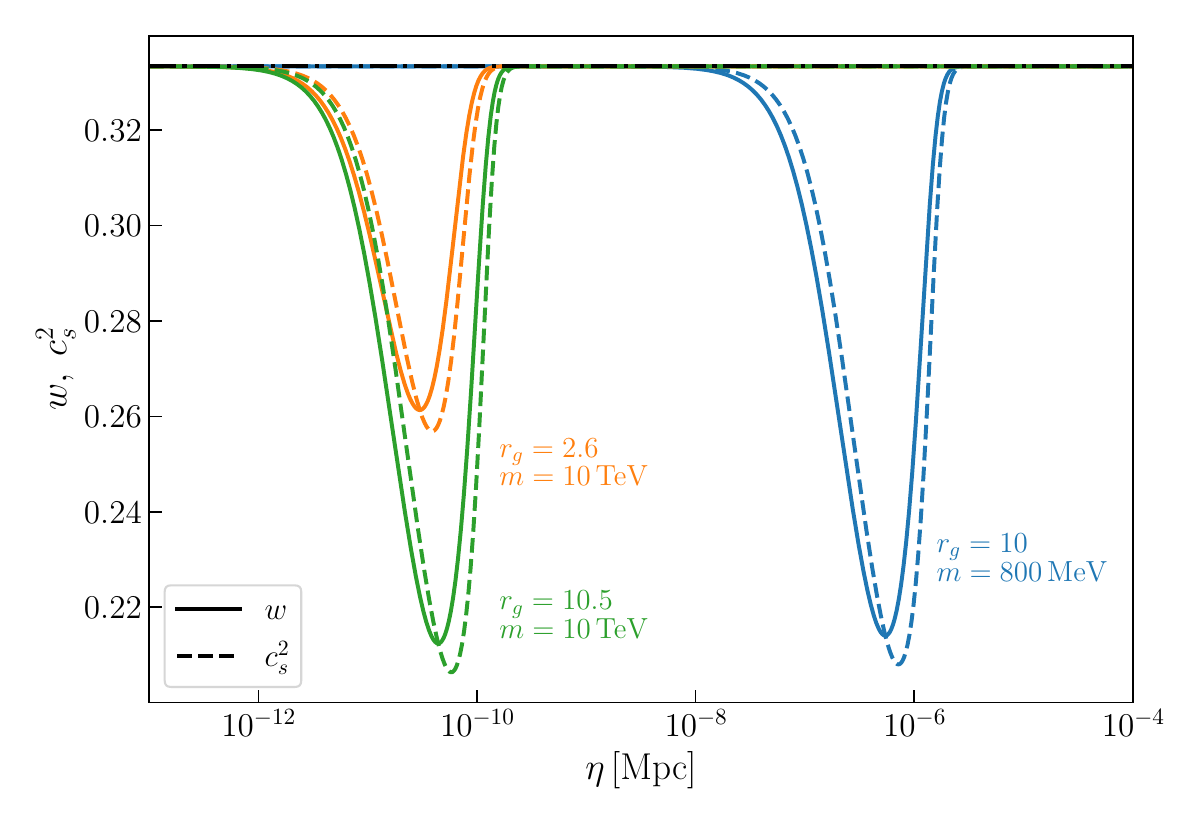}
    \caption{Equation-of-state parameter \(w\) and sound speed squared \(c_s^2\) in the toy model, for the benchmark values \((r_g=10,\,m=800\,\mathrm{MeV})\), \((r_g=2.6,\,m=10\,\mathrm{TeV})\), and \((r_g=10.5,\,m=10\,\mathrm{TeV})\). The dot-dashed line indicates the value for the standard radiation case ($w=c_s^2=1/3$). }
    \label{fig:Eos_model_draft}
\end{figure}

With the above expression, it is straightforward to see that the thermal history is controlled by two parameters $(m,r_g)$: the transition time is controlled by $T(\eta)\sim m$, while the depth of the suppression in $w$ and $c_s^2$ is controlled by $r_g$. 
We show examples of $w$ and $c_s^2$ for different choices of $m$ and $r_g$ in Fig.~\ref{fig:Eos_model_draft}. As seen from this figure, heavier masses lead to earlier transitions. Both \(w\) and \(c_s^2\) develop dips around the transition, with a slightly deeper and later minimum in \(c_s^2\)
than in \(w\). The depths of the dips are controlled by \(r_g\). As will be discussed later, these features are similar to those in real phase transitions in the SM. Here we take \((r_g,m)=(10,\,800\,\mathrm{MeV})\) to mimic the QCD crossover, and \((r_g,m)=(2.6,\,10\,\mathrm{TeV})\) and \((10.5,\,10\,\mathrm{TeV})\) for the composite-Higgs transition with \(N_c=5,10\) at \(T_c=10\,\mathrm{TeV}\) discussed in Sec.~\ref{subsec:bsm}.

\subsection{Phase Transitions in the Standard Model}
\label{sec:eos_sm_thermal}
The toy model only illustrates basic features of a single phase transition, but the thermal history of the SM is much richer.
Three main processes in the SM thermal history change the effective number of
relativistic degrees of freedom and hence drive \(w(\eta)\) and \(c_s^2(\eta)\) away from
\(1/3\). The electroweak crossover occurs at \(T\sim100\,\mathrm{GeV}\)~\cite{Englert:1964et,Higgs:1964pj,Weinberg:1967tq}.
As the plasma cools below this scale, the Higgs acquires its vacuum expectation value and \(W\), \(Z\), top quark and Higgs
contributions pass from relativistic to non-relativistic. The QCD crossover occurs at \(T\sim100\,\mathrm{MeV}\)~\cite{Borsanyi:2016,Bhattacharya:2014ara}. The strong interaction confines quarks and gluons into hadrons. Therefore, \(g_*\) drops from the quark--gluon value to the hadronic one during QCD crossover.
Finally, electron-positron annihilation takes place at \(T\sim500\,\mathrm{keV}\)~\cite{Saikawa:2018rcs}. Once the photon temperature falls below the electron mass, \(e^-e^+\) pairs annihilate into photons,
transferring their entropy to the photon bath.\footnote{Around the same time as \(e^-e^+\) annihilation, there is another important process called neutrino decoupling. Around neutrino decoupling, the fluid becomes non-ideal and has a dissipative effect on GWs~\cite{Domenech:2025bvr,Yu:2025cqu}. After neutrino decoupling, the single-temperature treatment becomes approximate. Free-streaming neutrinos would have additional effects on SIGW~\cite{Weinberg:2003ur,Saga:2014jca,Zhang:2022dgx}. Since the GW observations studied in this work are associated with higher scales such as QCD crossover, these effects are negligible and thus not considered in this work.} In each case the thermal bath temporarily
contains a mixture of relativistic and non-relativistic or intermediate species, which is
precisely the situation captured by \(w(\eta)\neq1/3\) and \(c_s^2(\eta)\neq w(\eta)\).

\begin{figure}[ht]
    \centering
    \includegraphics[width=0.65\linewidth]{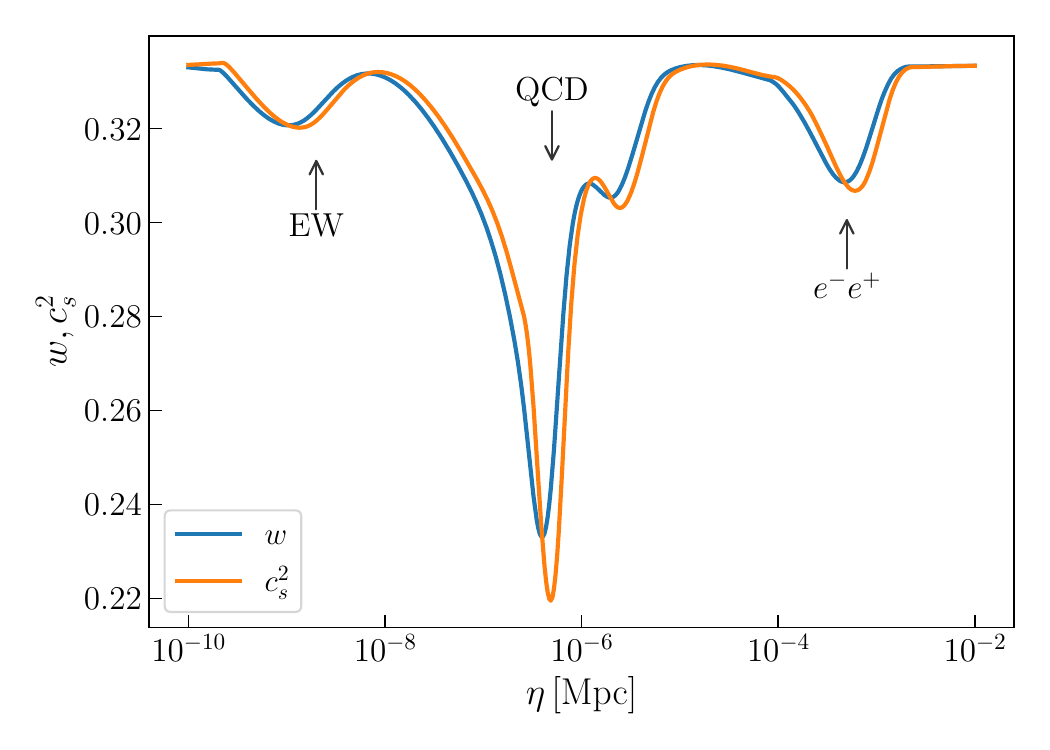}
    \caption{Equation-of-state \(w\) and sound speed squared \(c_s^2\) across the electroweak phase transition, lattice-QCD crossover and electron-positron annihilation.}
    \label{fig:Eos_QCD_draft}
\end{figure}

Since QCD crossover involves non-perturbative dynamics, the precise values of \(g_*\) and \(g_{*s}\) during this period require lattice QCD calculation. However, due to uncertainties in the lattice data and fitting functions, the resultant $w$ and $c_s^2$ used in existing SIGW studies do not conserve entropy, as also pointed out in Ref.~\cite{Abe:2020sqb}. Since the current formalism of the SIGW calculation is for a single adiabatic fluid with entropy conservation, there is a potential concern whether some features in SIGW spectrum are artifacts of entropy non-conservation. In this work, we present a systematic way of constructing \(g_*\) and \(g_{*s}\) with entropy conservation based on lattice QCD input. 
Instead of using both $g_*$ and $g_*/g_{*s}$ data from a lattice QCD calculation~\cite{Borsanyi:2016} to obtain interpolation functions of $g_*$ and $g_{*s}$, we only use values of \(g_*/g_{*s}\) and compute the corresponding \(w\) from Eq.~\eqref{eq:wcs_thermo}. The values of \(c_s^2\) are subsequently calculated from \(w\) through the adiabatic relation in Eq.~\eqref{eq:adiabatic_cs_w}, which ensures conservation of entropy. There is some freedom in interpolating the data, which introduces some ambiguity in \(w\) and \(c_s^2\). We emphasize that, for the entropy-conserving input constructed here, the reconstructed \(g_*\) and \(g_{*s}\) are within the uncertainty of the lattice QCD results. The details of the construction are described in Appendix~\ref{app:sm_eos}. Finally, we obtain the electron-positron annihilation curve from the standard relativistic thermodynamics and combine it with the reconstructed QCD input.
The resulting profiles are shown in Fig.~\ref{fig:Eos_QCD_draft}, which clearly shows three transitions in the SM: EW and QCD crossovers, as well as $e^-e^+$ annihilation. Each of the transitions exhibits a dip in both $w$ and $c_s^2$, similar to those of the toy model. The QCD crossover is more complicated, which has a small valley in addition to the main dip.

\subsection{Phase Transitions beyond the Standard Model}
\label{subsec:bsm}
BSM models that address the electroweak hierarchy problem often predict new dynamics not far from the weak scale. In this work, we consider the composite Higgs models that generally have two sectors: a strongly coupled composite sector with approximately conformal symmetry and an elementary sector with SM degrees of freedom that mixes with the composite sector. Below the critical temperature $T_c$, the composite sector undergoes a confinement phase transition, and Higgs bosons are born as fully composite bosonic states while other SM fields are a mixture of elementary and composite states (known as partial compositeness). Since this kind of phase transition involves non-perturbative dynamics, it is technically challenging to have a quantitative study in four dimensions. According to AdS/CFT, four-dimensional strongly coupled composite Higgs theories are dual to weakly coupled five-dimensional Randall-Sundrum models. We use the 5D dual models to fix the order one factors in this study.  

To study the impact of confinement phase transition associated with composite Higgs models, we use the toy model in Sec.~\ref{sec:toy_model} to model the change of $w$ and $c_s$ during this phase transition.\footnote{As argued in Refs.~\cite{Creminelli:2001th}, this type of confinement phase is typically first-order, meaning there is generally some entropy injection to the SM bath via the latent heat. However, as studied in Ref.~\cite{Agashe:2019lhy,Agashe:2020lfz}, such transition can become prompt near $T_c$ and thus the latent heat injection can be negligible. We therefore assume the entropy is conserved during such phase transition and apply our method to calculate SIGW in this work.}
The relativistic degrees of freedom are the sum of the composite and elementary contributions: \(g_{*,\rm UV}=g_{*,\rm comp}+g_{*,\rm elem}\). The free energy of the deconfined composite sector with \(SU(N_c)\) gauge symmetry is \(F_{\rm comp}=-\frac{\pi^2}{8}N_c^2 T^4\) for large \(N_c\), with the numerical factor fixed using the 5D dual models~\cite{Creminelli:2001th,Bruggisser:2018mrt}. This result is consistent with a thermal radiation sector with \(O(N_c^2)\) degrees of freedom. With the definition of the free energy density, we can obtain $g_{*,\rm comp}$ as 
\begin{equation}
    g_{*,\rm comp}=\frac{45}{4}N_c^2.
\end{equation}
The elementary sector contains all SM fields except the Higgs doublet: $g_{*,\rm elem}=g_{*,\rm SM}-4=102.75$. After the phase transition, only the SM degrees of freedom remain: \(g_{*,\rm IR}=g_{*,\rm SM}\). We can therefore obtain the ratio \(r_g\) as
\begin{equation}
    r_g=\frac{g_{*,\rm UV}-g_{*,\rm IR}}{g_{*,\rm IR}}=\frac{\frac{45}{4}N_c^2-4}{106.75}.
\end{equation}

The shapes of $w$ and $c_s^2$ of composite Higgs models are shown in Fig.~\ref{fig:Eos_model_draft}. Here we choose the critical temperature $T_c=10$ TeV and consider two different $N_c$: $N_c=5$ and $N_c=10$. Therefore, these two composite Higgs cases can be mapped to $(r_g=2.6,m=10\,\textrm{TeV})$ and $(r_g=10.5,m=10\,\textrm{TeV})$ in the toy model.

\section{Scalar Induced Gravitational Waves from Phase Transitions}
\label{sec:thermal}

In this section, we calculate SIGWs based on $w$ and $c_s^2$ obtained from the previous section for different models. To clearly view the features of induced GW spectrum from a single phase transition, we first study the toy model with a flat curvature power spectrum. We provide physical explanations of those features. We then show results for the SM and BSM cases. Moreover, we present GW spectra from delta-function and broken power law curvature spectra.
\subsection{Scale-invariant Curvature Power Spectrum}
\label{sec:flat spectrum}

\subsubsection{Toy Model}

\begin{figure}[ht]
    \centering
\includegraphics[width=0.7\linewidth]{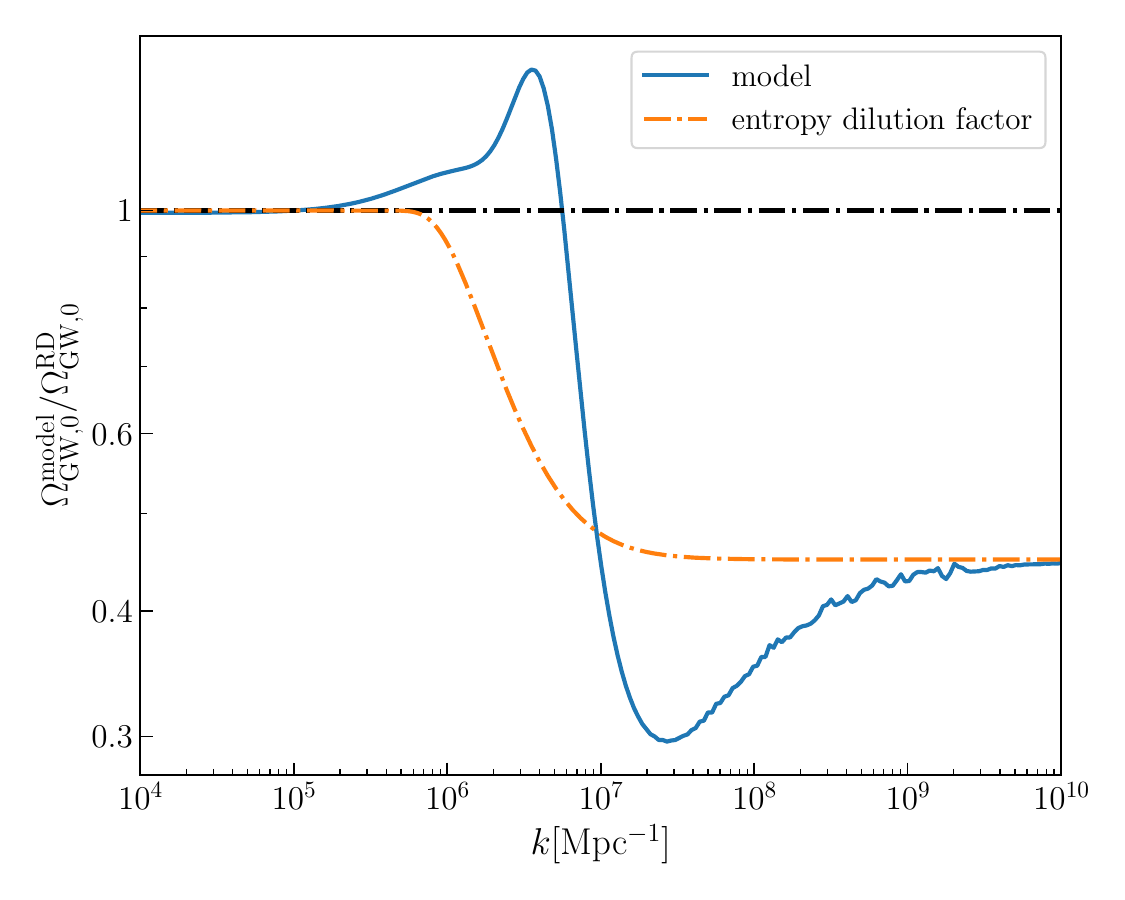}
    \caption{SIGW spectrum in the toy model with ($r_g=10,m=800\,\textrm{MeV}$), normalized to the SIGW spectrum in the standard radiation case. For both cases, we use the same scale-invariant primordial curvature power spectrum $\mathcal P_\zeta=A$. The orange dot-dashed line shows the entropy dilution factor $R(k)$ with \(\eta_{\rm ref}=10^{-4}\,\mathrm{Mpc}\).}
    \label{fig:flat_model_draft}
\end{figure}

To clearly study the features of the SIGW spectrum from a single phase transition, we apply our toy model discussed in Sec.~\ref{sec:toy_model} and choose a flat primordial curvature power spectrum $\mathcal P_{\zeta}=A$. Applying the formalism in Sec.~\ref{sec:formalism}, we obtain the GW spectrum in Fig.~\ref{fig:flat_model_draft} with parameters ($r_g=10,m=800\,\textrm{MeV}$) of the toy model. As seen in Fig.~\ref{fig:flat_model_draft}, the most prominent feature in the SIGW spectrum from a phase transition is the step-like shape. The size of the step is controlled by the entropy dilution. The transition frequency depends on the start of the phase transition, which is larger than the transition scale of the entropy dilution. A peak and a dip around the transition are due to the values of $c_s^2$ and $w$.
\paragraph{Entropy Dilution Factor.} Fig.~\ref{fig:flat_model_draft} shows two approximately flat plateaus in both the ultraviolet (UV) and infrared (IR) regions. The UV plateau is lower because larger \(k\) modes are generated earlier, where they experience a stronger suppression due to the heating of the photon bath from the change of $g_{*s}$. We can estimate the size of this suppression for a given $k$ mode as the entropy dilution factor \(R\) between a time $\eta=1/k$ and a reference time $\eta_{\rm ref}$ in the IR plateau 
\begin{equation}
    R(k)=\frac{g_{*}(1/k)}{g_{*}(\eta_{\rm ref})}
    \left(\frac{g_{*s}(\eta_{\rm ref})}{g_{*s}(1/k)}\right)^{4/3} .
    \label{eq:R_draft}
\end{equation}
As shown in Fig.~\ref{fig:flat_model_draft}, $R(k)$ smoothly connects two plateaus and the ratio of two plateaus nicely matches the expected value
\begin{equation}
 \frac{\Omega_{\rm GW,0}^{\rm UV}}{\Omega_{\rm GW,0}^{\rm IR}}
 \simeq\left(\frac{g_{*,\rm UV}}{g_{*,\rm IR}}\right)^{-1/3}
 =(1+r_g)^{-1/3}\simeq0.449\quad(r_g=10).
 \label{eq:plateau_ratio}
\end{equation}

\paragraph{Transition Frequency.} We can estimate the transition frequency as 
\begin{equation}
\label{eq:f_tr}
    f_{\rm tr}\sim \frac{k_{\rm start}}{2\pi},
\end{equation}
where $k_{\rm start}$ is the comoving wavenumber at the start of the change of $w$ and $c_s^2$. For the case shown in Fig.~\ref{fig:flat_model_draft}, $k_{\rm start}\sim 10^{7}\,\textrm{Mpc}^{-1}$ or $f_{\rm tr}\sim 10^{8}\,\textrm{Hz}$. We note that this $k_{\rm start}$ is larger than the transition scale of the entropy dilution factor, which is typically determined by the scale where $w$ and $c_s^2$ reach the minimum. Therefore, the transition scale in GW spectrum is shifted towards a higher value compared to the entropy dilution factor shown in Fig.~\ref{fig:flat_model_draft}. For a general case, we can first identify the $k_{\rm start}$ based on the shape of $w$ and $c_s^2$, and then use Eq.~\eqref{eq:f_tr} to estimate the transition frequency.

\begin{figure}[htbp]
    \centering
    \includegraphics[width=0.48\linewidth]{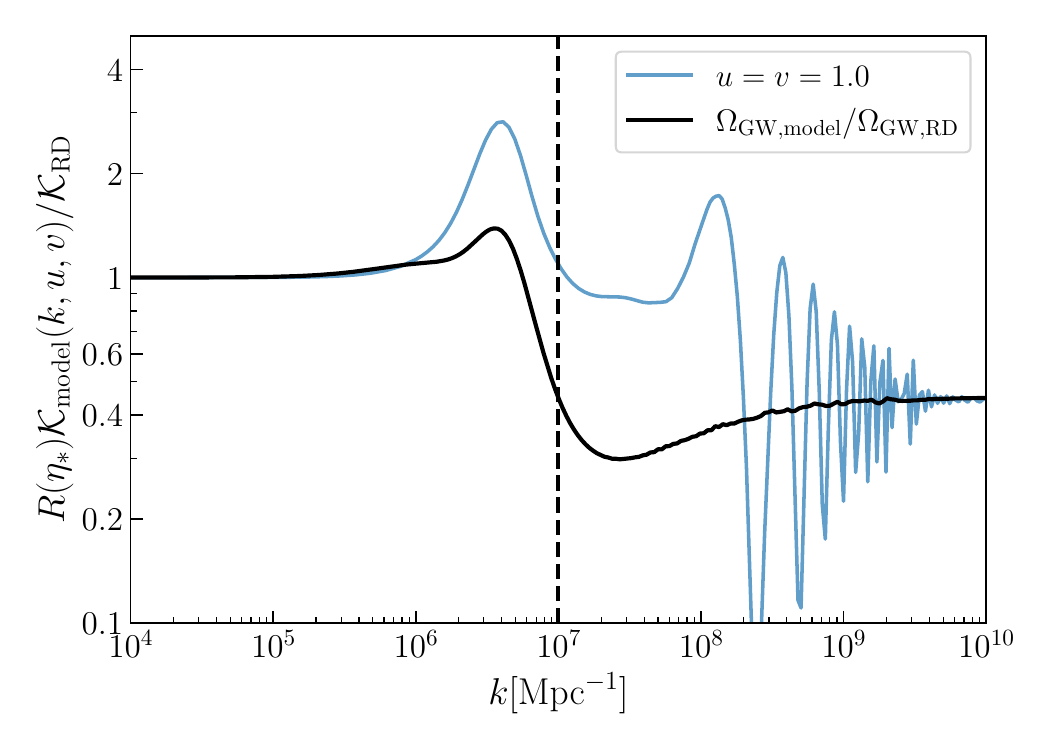}\hfill
    \includegraphics[width=0.48\linewidth]{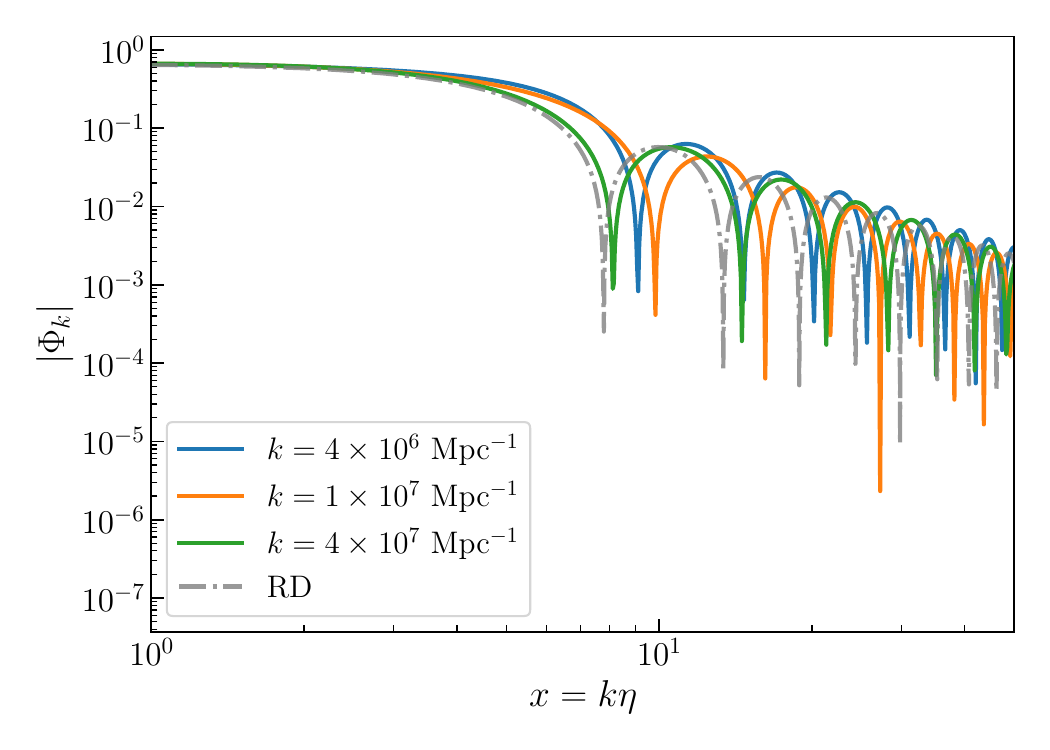}
    \caption{\textit{Left}: the ratio \(R(\eta_*){\cal K}_{\rm model}(k,u,v)/{\cal K}_{\rm RD}\) for \(u=v=1\), compared with the ratio \(\Omega_{\rm GW,0}^{\rm model}/\Omega_{\rm GW,0}^{\rm RD}\). The vertical dashed line marks \(k=10^7\,\mathrm{Mpc}^{-1}\). \textit{Right}: evolution of the scalar transfer function in the toy model, compared with the standard radiation solution, with \(x\equiv k\eta\). }
    \label{fig:model_Phi_draft}
\end{figure}

\paragraph{A Peak and a Dip.}
In addition to the step, Fig.~\ref{fig:flat_model_draft} exhibits a peak on the
IR side and a dip on the UV side of the transition region.
These local extrema are distinct features associated with phase transitions. Their shapes reflect the combined effects of the time-dependent background with $c_s^2<1/3$ and $w<1/3$.

We start with the discussion on the origin of the peak on the IR side. For $k\lesssim k_{\rm start}\sim 10^{7}\,\textrm{Mpc}^{-1}$, these modes enter the sound horizon during or after the transition. As shown in Fig.~\ref{fig:model_Phi_draft}, scalar perturbations $\Phi_k$ start to damp and oscillate once they enter the sound horizon when $c_sk \eta \sim 1 $. Since $c_s^2$ drops below the standard radiation value $1/3$ during the phase transition, the expected horizon entry is delayed, resulting in a relatively enhanced amplitude compared to the standard radiation case. The maximum enhancement occurs when the mode enters the sound horizon at the time $\eta_{\rm min}$ where $c_s^2$ reaches the minimum: $c_{s,\rm min}k_{\rm peak}\eta_{\rm min}\sim1$. For the case shown in Fig.~\ref{fig:flat_model_draft}, $\eta_{\rm min}\sim 7\times 10^{-7}\,\textrm{Mpc}$ and thus $k_{\rm peak}\sim 4\times 10^{6}\,\textrm{Mpc}^{-1}$. Enhanced scalar perturbations lead to an enhanced source kernel $\mathcal K$.\footnote{We define the source kernel $\mathcal K$ as
\begin{equation}
   {\cal K}(k,u,v) = \l(\frac{k}{\mathcal H(\eta_*)}\r)^2\overline{ I^2(k,u,v)}~,
\end{equation}
where \(I^2(k,u,v)\equiv I^2(k,uk,vk,\eta_*)\). 
Therefore, this kernel is related to the final spectrum as
\[
    \d\Omega_{\rm GW,0}^{\rm model}/\d u\,\d v\propto R(\eta_*)\,{\cal K}(k,u,v) ,
\]
where \(R(\eta_*)\) is the entropy dilution factor.} As shown in Fig.~\ref{fig:model_Phi_draft}, the source kernel is enhanced for modes entering the sound horizon during the transition, with a peak at $k_{\rm peak}$, agreeing with our previous estimation.

For \(k\gtrsim k_{\rm start}\sim10^7\,\mathrm{Mpc}^{-1}\), the phase transition happens inside the oscillation region of \(\Phi_k\), so we can solve the subhorizon solution with the WKB method. Deep inside the sound horizon, 
\begin{equation}
    \Phi_k(\eta)\sim C_k\exp\left[-\frac32\int^\eta\mathcal H(1+c_s^2)\,d\tilde\eta\right]\exp\left[\pm ik\int^\eta c_s(\tilde\eta)d\tilde\eta\right],
    \label{eq:wkb_ansatz_draft}
\end{equation}
which reduces to the radiation solution \(\Phi_k=C_k\eta^{-2}\exp(\pm ic_sk\eta)\) for constant \(c_s^2=1/3\) and \(\mathcal H=1/\eta\). Here \(C_k\) is a constant fixed by matching at the starting point of the WKB method. With Eq.~\eqref{eq:wkb_ansatz_draft}, we can understand that the amplitude (the first exponential) is reduced, relative to the radiation case, for $w<1/3$ because the Hubble parameter is enhanced. The phase of the oscillation is determined by the $k\int c_s d\eta$, which leads to a phase delay due to $c_s<1/\sqrt{3}$ compared to the standard radiation case. Both features are reflected in scalar perturbations shown in the right panel of Fig.~\ref{fig:model_Phi_draft}. The phase difference averages out in the final GW spectrum, leaving the net amplitude suppression and thus a dip on UV side of the step.

To summarize, the rise of the spectrum on the IR side of the transition is due to $c_s^2<1/3$, with the peak scale satisfying $c_{s,\rm min}k_{\rm peak}\eta_{\rm min}\sim 1$. The suppression on the UV side is due to $w<1/3$ that reduces the amplitude of sub-horizon scalar perturbations. We also perform a few test calculations shown in Appendix~\ref{sec:numerical check}, which justifies our previous estimates.

\subsubsection{Phase Transitions in the Standard Model and Beyond}
\label{sec:sm_thermal}

\begin{figure}[t]
    \centering
    \includegraphics[width=0.8\linewidth]{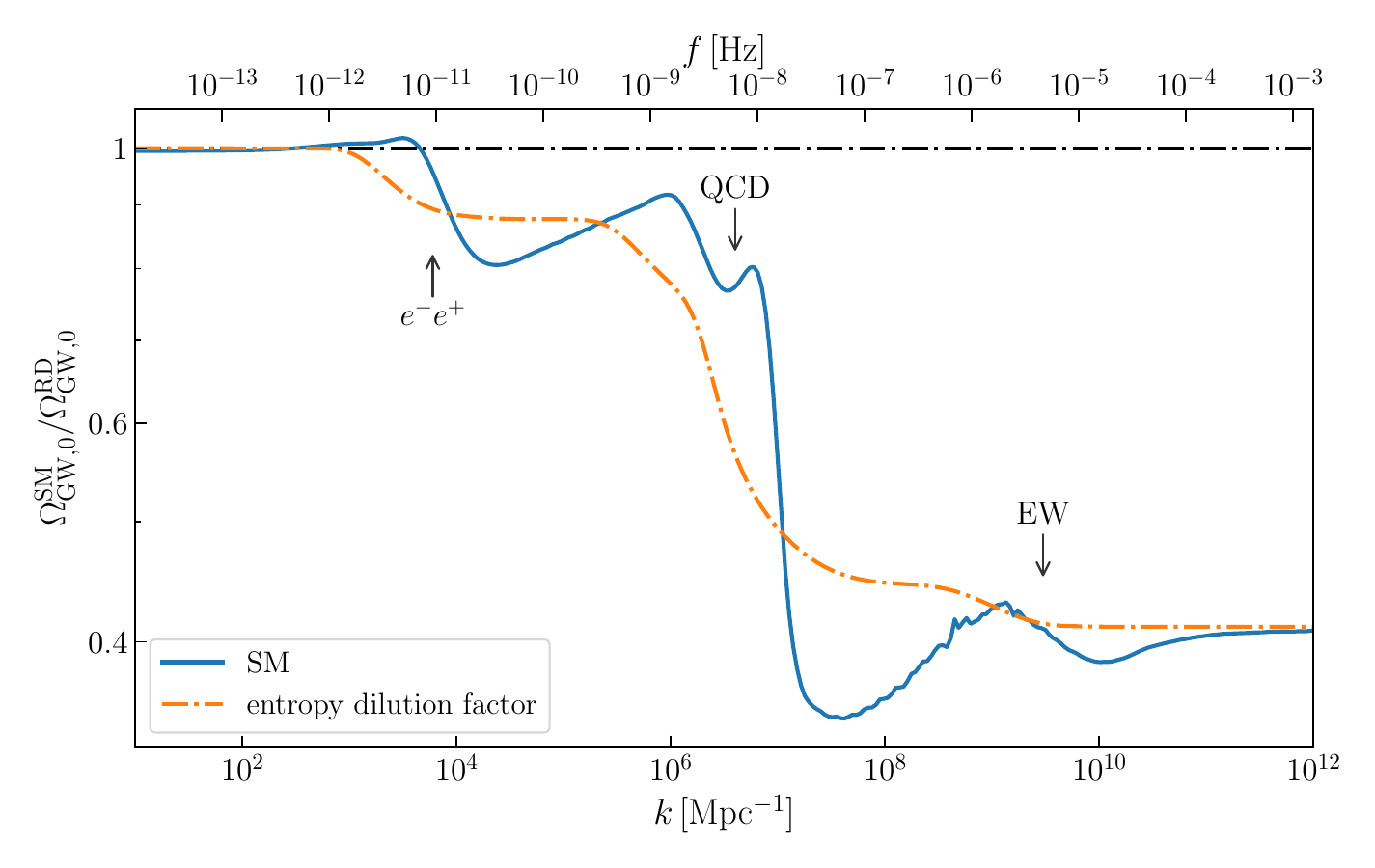}
    \caption{SIGW spectrum from phase transitions in the Standard Model with $\mathcal P_\zeta(k)=A$, normalized to the GW spectrum in the standard radiation case. The orange dot-dashed line shows the entropy dilution factor $R(k)$ with \(\eta_{\rm ref}=10^{-2}\,\mathrm{Mpc}\).}
    \label{fig:flat_QCD_draft}
\end{figure}

With the input of SM thermal history from Sec.~\ref{sec:eos_sm_thermal}, the induced GW spectrum from a flat curvature spectrum is shown in Fig.~\ref{fig:flat_QCD_draft}. This GW spectrum clearly shows three transitions in the SM, each of them having the step-like feature discussed in the toy model. On the
left, around \(k\sim10^{4}\,\mathrm{Mpc}^{-1}\)($f\sim10^{-11}\,\textrm{Hz}$), the step and the nearby
small peak--dip structure are generated by electron-positron annihilation. Around
\(k\sim10^{7}\,\mathrm{Mpc}^{-1}\)($f\sim10^{-8}\,\textrm{Hz}$), the larger step and the sequence of large and
small extrema are associated with the QCD crossover. This special feature is because  
\(w(\eta)\) and \(c_s^2(\eta)\) during QCD epoch have a small dip in addition to the main dip, whose responses overlap in the induced spectrum.
Finally, the feature around \(k\sim10^{9}\,\mathrm{Mpc}^{-1}\)($f\sim10^{-6}\,\textrm{Hz}$) is caused by the
electroweak crossover. The full spectrum exhibits combined effects from each transition, although
individual ones are not cleanly resolved because their $w$ and $c_s^2$ overlap.

The entropy dilution factor defined in Eq.~\eqref{eq:R_draft} for SM phase transitions is also shown in Fig.~\ref{fig:flat_QCD_draft}. It nicely matches the SIGW spectrum at both UV and IR plateaus, with the plateau ratio 
\(\Omega_{\rm GW,0}^{\rm UV}/\Omega_{\rm GW,0}^{\rm IR}\simeq 0.412\). The steps in the dilution factor also indicate the scales for each transition.
For the same scale-invariant primordial spectrum, the full SM calculation in
Fig.~\ref{fig:flat_QCD_draft} gives a maximum suppression of about 65\% relative to the standard radiation case in the PTA band, with \(\Omega_{\rm GW,0}^{\rm SM}/\Omega_{\rm GW,0}^{\rm RD}\simeq0.35\)
near \(f\simeq46\,\mathrm{nHz}\).
This includes the change in GW production as well as entropy dilution.

\begin{figure}[h]
    \centering
    \includegraphics[width=0.8\linewidth]{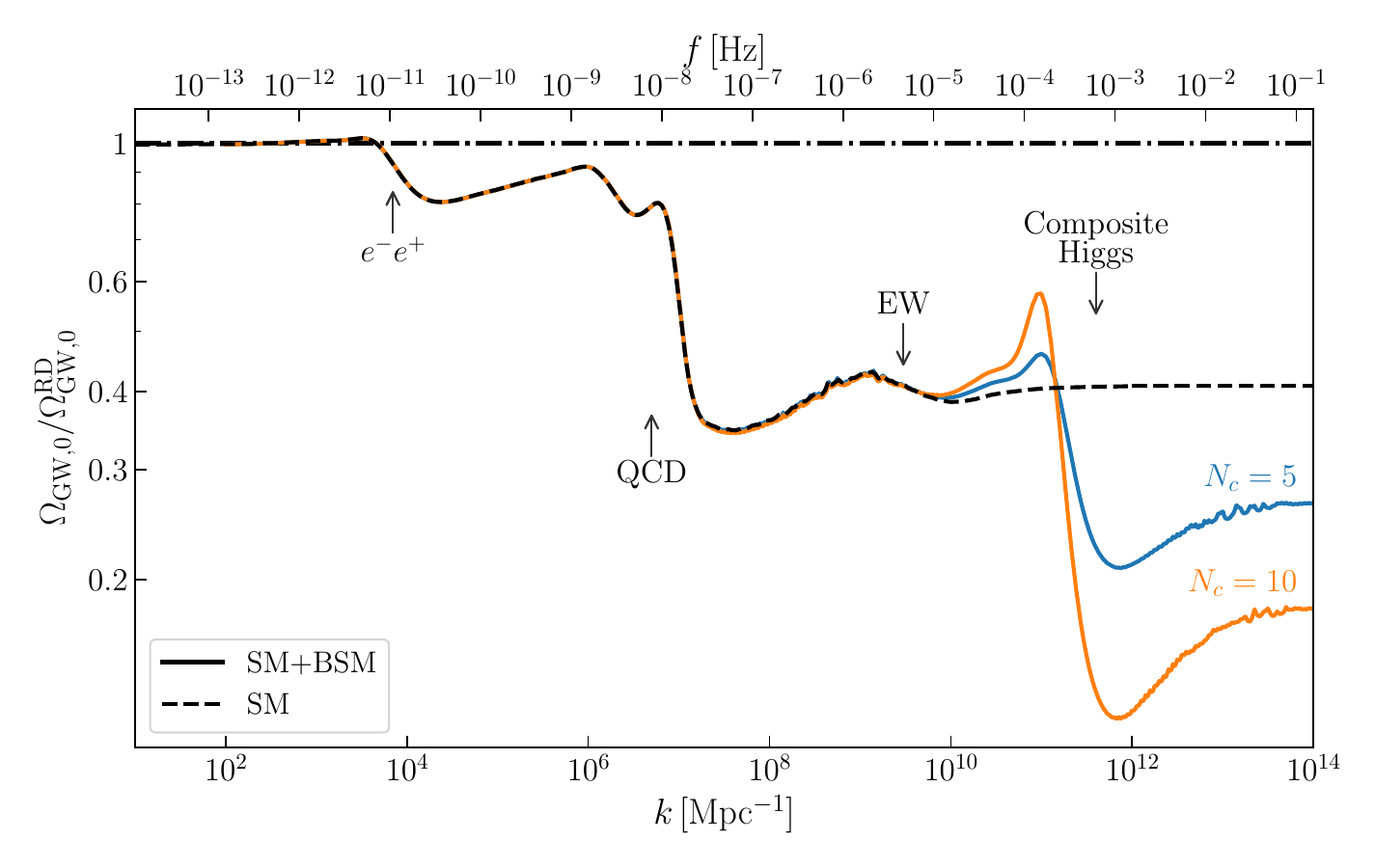}
    \caption{SIGW spectrum from phase transitions in the SM and the composite Higgs model with $T_c=10\,\mathrm{TeV}$,
    normalized to the standard radiation result. The blue and orange solid curves
    show the cases $N_c=5$ and $N_c=10$, respectively. The black dashed curve
    gives the SM result without the composite sector.}
    \label{fig:flat_BSM}
\end{figure}

As an example of a BSM case, we choose to study phase transitions in composite Higgs models. We apply $w$ and $c_s^2$ discussed in Sec.~\ref{subsec:bsm} for the critical temperature $T_c=10\,\mathrm{TeV}$ and the two benchmark choices $N_c=5$ and $10$. The induced GW spectrum is shown in
Fig.~\ref{fig:flat_BSM}.
The confinement phase transition in the composite Higgs model produces a step at $f\sim 10^3\,\rm Hz$ and lowers the UV plateau due to entropy dilution. This suppression is stronger for $N_c=10$ (a $\sim 60\%$ difference between UV and IR plateaus), because it has larger degrees of freedom in the composite sector and hence a larger $r_g$.

\begin{figure}[ht]
    \centering
    \includegraphics[width=0.48\linewidth]{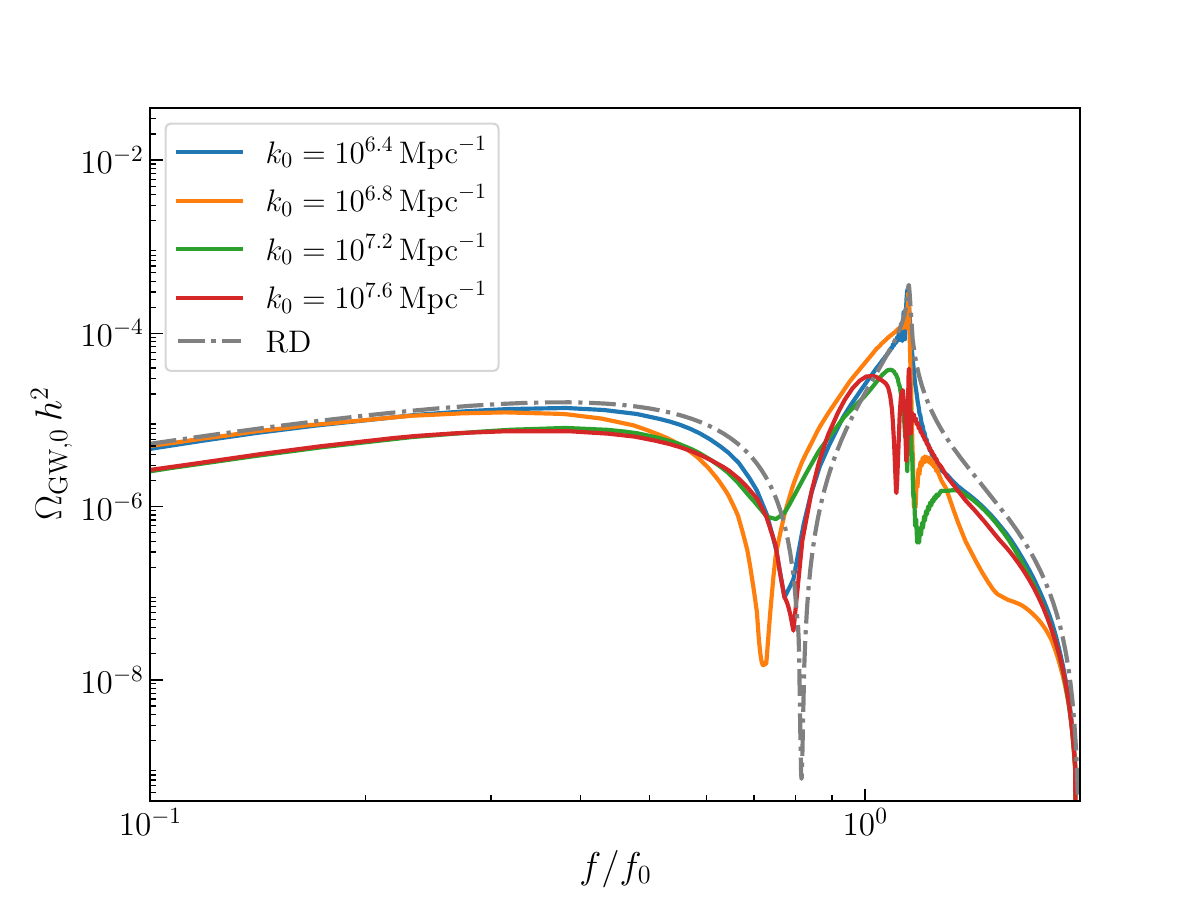}
    \includegraphics[width=0.48\linewidth]{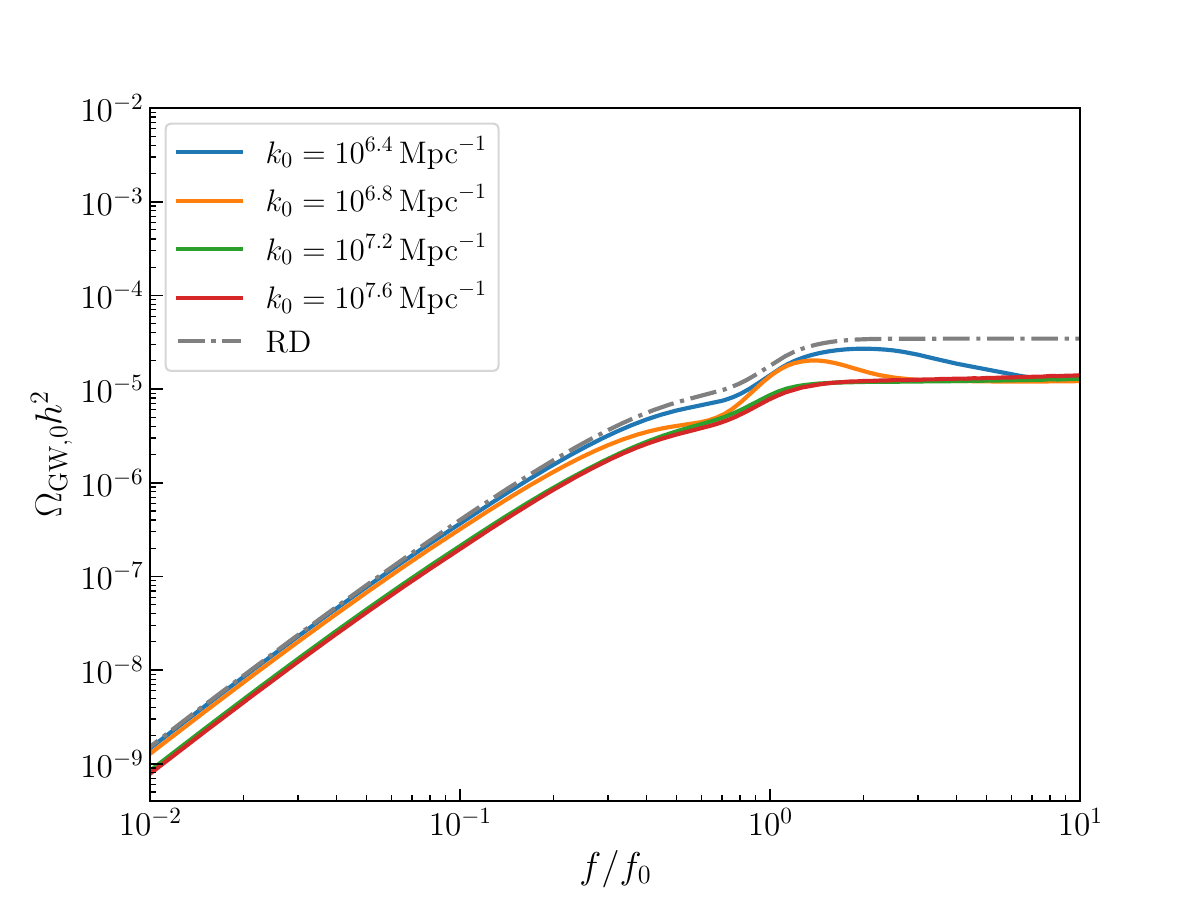}
    \caption{Present-day SIGW spectra in the SM thermal history sourced by a delta-function spectrum (left) and a BPL spectrum (right) for several wavenumbers around the QCD crossover, together with the standard radiation prediction. Here we choose the amplitude of the curvature spectrum $A=1$.}
\label{fig:delta_BPL_QCD_draft}
\end{figure}

\subsection{Delta-function and Broken Power Law Curvature Spectra}
In this section, we present SIGW spectra for the delta-function and BPL curvature spectra in Fig.~\ref{fig:delta_BPL_QCD_draft}. Here we focus on cases of the SM thermal history, which are relevant for data analysis in the next section. In the left panel of Fig.~\ref{fig:delta_BPL_QCD_draft}, we show SIGWs for a delta-function spectrum (Eq.~\eqref{eq:P_delta_draft}) with four different $k_0$ around the QCD crossover. The four values of \(k_0\) are chosen such that their horizon entry spans from before to after the QCD sound-speed variation. The SIGW spectrum in the standard radiation case (the dot-dashed line) has a resonance peak at $k=2c_s k_0$ and cancellation valley at $k=\sqrt{2}c_s k_0$. The positions of the peak and valley have been demonstrated to be valid for general constant values of $w=c_s^2$~\cite{Domenech:2019quo}. For a time-dependent background, it has been shown in Ref.~\cite{Abe:2020sqb} that the cancellation valley still follows the similar scaling $k=\sqrt{2}c_s k_0$, but the value of $c_s$ should be chosen at $\eta\sim 4/k_0$. Our result for the SM background in Fig.~\ref{fig:delta_BPL_QCD_draft}
is roughly consistent with this scaling. 
The resonance peak, however, does not exhibit a simple scaling. For modes entering the horizon before or around the early stage of the QCD crossover, the single resonance peak of the radiation case splits into multiple small peaks, which will be smeared out for a broader spectrum (e.g., log-normal spectrum).  
Moreover, a new broad peak developed near $k\sim k_0$. The same features have also been shown in Ref.~\cite{Feng:2026xwp}. In the far infrared, all GW spectra have the universal slope \(\Omega_{\rm GW}\propto k^2\ln^2(k_0/k)\)~\cite{Yuan:2019wwo}.

In the right panel of Fig.~\ref{fig:delta_BPL_QCD_draft}, we show results for the BPL spectrum in Eq.~\eqref{eq:P_BPL_draft} with four
turnover scales ($k_0$) around the QCD crossover. All curves retain the expected infrared scaling \(\Omega_{\rm GW}\propto k^3\ln^2(k_0/k)\)~\cite{Yuan:2019wwo}. However,
the results in the SM background approach lower UV plateaus than the standard radiation curve, reflecting
the entropy dilution. Since the BPL spectrum is scale-invariant above \(k_0\), its UV branch should be consistent with the flat-spectrum result in Fig.~\ref{fig:flat_QCD_draft}. The features around $f\sim f_0$ are more complicated, which could be used to distinguish a time-dependent background from the standard radiation case.

\section{Probing Phase Transitions with Gravitational Wave Experiments}
\label{sec:pta}
In this section, we first show that current and future GW experiments can probe phase transitions in the SM and in BSM models. In particular, the effects from the QCD crossover lie in the frequency window of PTAs. We then apply current PTA data from NANOGrav to place constraints on SIGW from a delta-function and a BPL curvature spectrum.

\subsection{Current and Future Experiments}

\begin{figure}[h]
    \centering
    \includegraphics[width=0.8\linewidth]{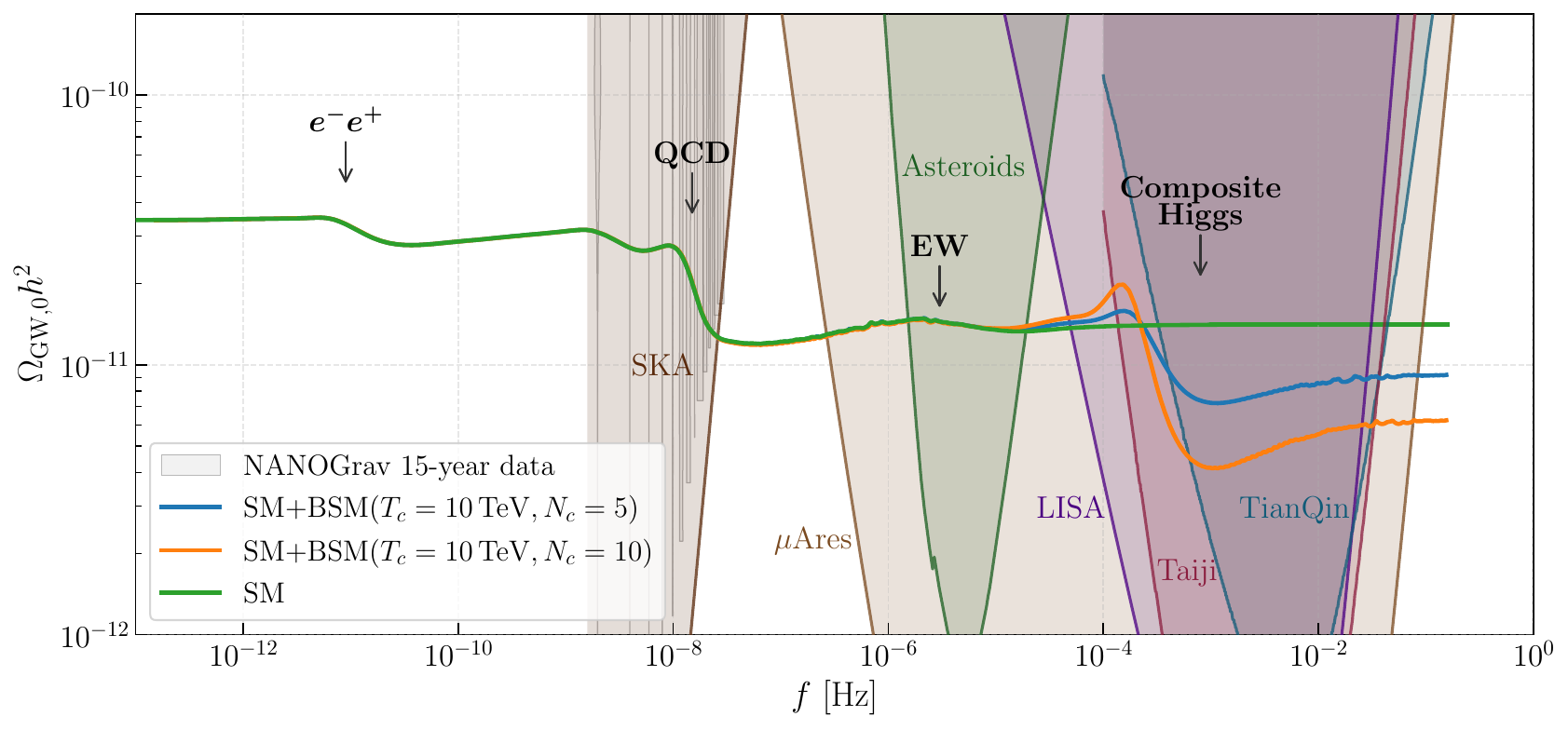}
    \caption{Present-day SIGW spectra with a flat curvature spectrum $\mathcal P_\zeta=10^{-3}$ for the combined Standard Model and composite-Higgs thermal history. We set the phase transition in the composite Higgs model to occur around \(T_c=10\,\mathrm{TeV}\), and we consider two values of \(N_c\): $N_c=5$ (blue) and $N_c=10$ (orange). The green solid line indicates the result with only the SM. The NANOGrav 15-year data are shown in gray, while the sensitivities of LISA~\cite{LISA:2017pwj,Baker:2019nia}, Taiji~\cite{Hu:2017mde,Ruan:2018tsw}, TianQin~\cite{TianQin:2015yph}, $\mu$Ares~\cite{Sesana:2019vho}, asteroid-based detectors~\cite{Fedderke:2021kuy} and SKA~\cite{Janssen:2014dka} are shown as shaded regions.}
    \label{fig:NANOGrav_flat_draft}
\end{figure}

Current and future GW observations across a wide range of frequencies are powerful probes of phase transitions in the SM and in BSM models. To demonstrate this, we show the SIGW spectrum for a scale-invariant curvature spectrum $\mathcal P_\zeta=10^{-3}$ in Fig.~\ref{fig:NANOGrav_flat_draft},\footnote{We note that the scale-invariant curvature spectrum  $\mathcal P _\zeta=10^{-3}$  at $k\lesssim 10^4\,\textrm{Mpc}^{-1}$ (or GW frequency $f\lesssim 10^{-11}\textrm{Hz}$) is ruled out by astrophysical and cosmological observations~\cite{10.1111/j.1365-2966.2011.18245.x,Cyr:2023pgw,Graham:2024hah,Qin:2025ymc}. One way to avoid such constraints is to suppress curvature spectrum for $k\lesssim 10^4\,\textrm{Mpc}^{-1}$ (see e.g.,~\cite{Buckley:2024nen,Barir:2022kzo, Kumar:2024hsi}). This means GW signals associated with $e^-e^+$ annihilation will also be modified accordingly. } together with current GW data and expected sensitivities of future GW experiments. As mentioned in Sec.~\ref{sec:sm_thermal}, the most prominent feature of phase transitions in the SM is the step associated with the QCD crossover, which occurs around $10^{-8}\,\textrm{Hz}$ and has a $\sim60\%$ difference between IR and UV plateaus. Fig.~\ref{fig:NANOGrav_flat_draft} clearly
shows that this transition lies within the current PTA observation window. Moreover, the SKA experiment can be a complementary probe of the QCD crossover. The signal of the EW crossover is a relatively small step transition around $\mu\textrm{Hz}$, which can be probed by future $\mu\textrm{Hz}$ GW observations, such as $\mu\textrm{Ares}$, asteroid-based experiments and GUEST~\cite{Blas:2026xol}. As an example of BSM physics, we show a phase transition in composite Higgs models around $T_c=10\,\textrm{TeV}$, which corresponds to a step transition around mHz. Future space-based experiments, such as LISA, Taiji and TianQin, cover this frequency window and thus are powerful probes of BSM phase transitions around TeV-PeV scales. Moreover, combined searches of current and future GW observations from nHz to mHz can reveal correlations between signals from phase transitions in the SM and BSM models, which can be used to greatly suppress noise.

\subsection{Constraints from Pulsar Timing Arrays}\label{sec:pta_constraint}

\begin{figure}[h]
    \centering
    \includegraphics[width=0.45\linewidth]{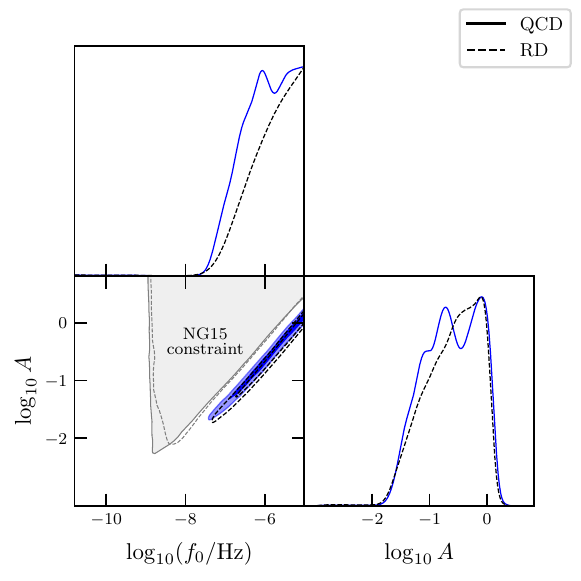}
    \includegraphics[width=0.45\linewidth]{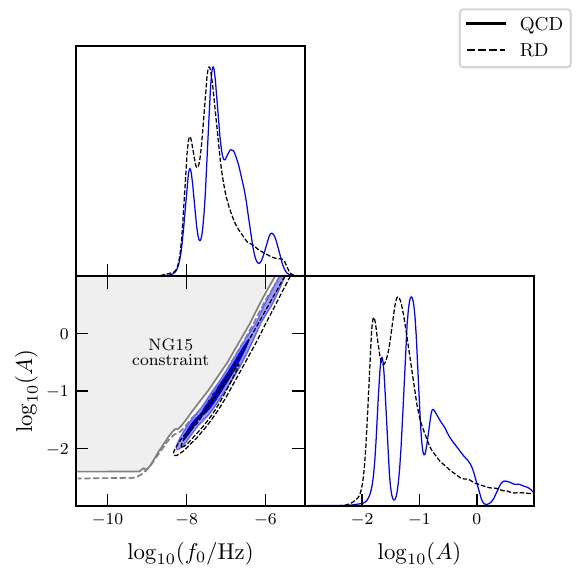}
    \caption{Best fits and constraints from the NANOGrav 15-year data for the delta-function (left) and broken-power-law (right) curvature spectra. Solid and dashed contours denote the results for the QCD crossover and the standard radiation case.}
    \label{fig:post_draft}
\end{figure}

Having discussed the shape of the SIGW spectra in Sec.~\ref{sec:sm_thermal}, we now fit the
NANOGrav 15-year data set~\cite{NANOGrav:2023gor,NANOGrav:2023hde,NANOGrav:2023hvm,nano_grav_2023_dataset} using the delta-function and BPL spectra for both the QCD crossover and the standard radiation case. 
For our data analysis, we
employ the peak or turnover frequency
\(f_0=k_0/(2\pi)\) and power spectrum amplitude \(A\) as parameters for each spectrum. We apply the
Bayesian inference method to determine the best fit of the SIGW spectra. We adopt \texttt{PTArcade}~\cite{andrea_mitridate_2023,Mitridate:2023oar} to
sample the posterior probability and generate upper limits above which the additional model is
``strongly disfavored'' according to the Jeffreys scale~\cite{NANOGrav:2023hvm,Jeffreys}. The priors on the parameters are log-uniform within
\(\log_{10}(f_0/\mathrm{Hz})\in[-10.8,-5]\) and \(\log_{10}A\in[-3,1]\). 
The resulting posterior distributions are shown in
Fig.~\ref{fig:post_draft}. The best-fit values for both backgrounds and the Bayes factors
are listed in Table~\ref{tab:PTA_results_draft}.

\begin{table}[h]
    \centering
    \begin{tabular}{llccc}
    \toprule
    Spectrum & Background & $\log_{10}(f_0/\mathrm{Hz})$ & $\log_{10}A$ & $\mathcal{B}_{\rm QCD/RD}$ \\
    \midrule
    Delta function       & QCD crossover & $-6.084$ & $-0.718$ & $1.06$ \\
    Delta function       & Standard RD & $-5.632$ & $-0.465$ & --- \\
    Broken power law & QCD crossover & $-7.301$ & $-1.097$ & $1.36$ \\
    Broken power law & Standard RD & $-7.388$ & $-1.382$ & --- \\
    \bottomrule
    \end{tabular}
    \caption{Best-fit frequency and amplitude for the QCD crossover and the standard radiation case.
    The Bayes factor $\mathcal B_{\rm QCD/RD}$ compares the QCD crossover and standard radiation cases.}
    \label{tab:PTA_results_draft}
\end{table}

\begin{figure}[h]
    \centering
    \includegraphics[width=0.8\linewidth]{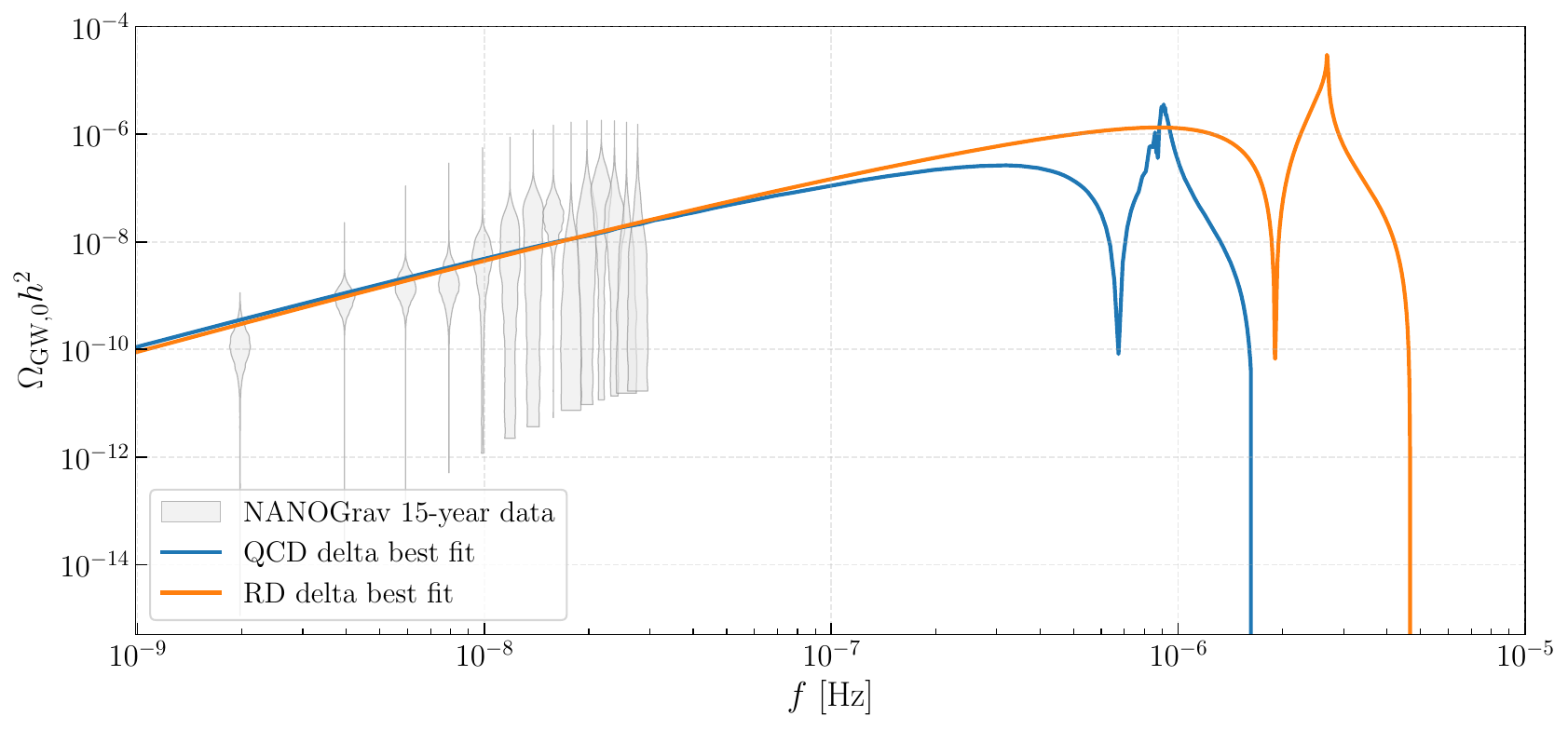}
    \includegraphics[width=0.8\linewidth]{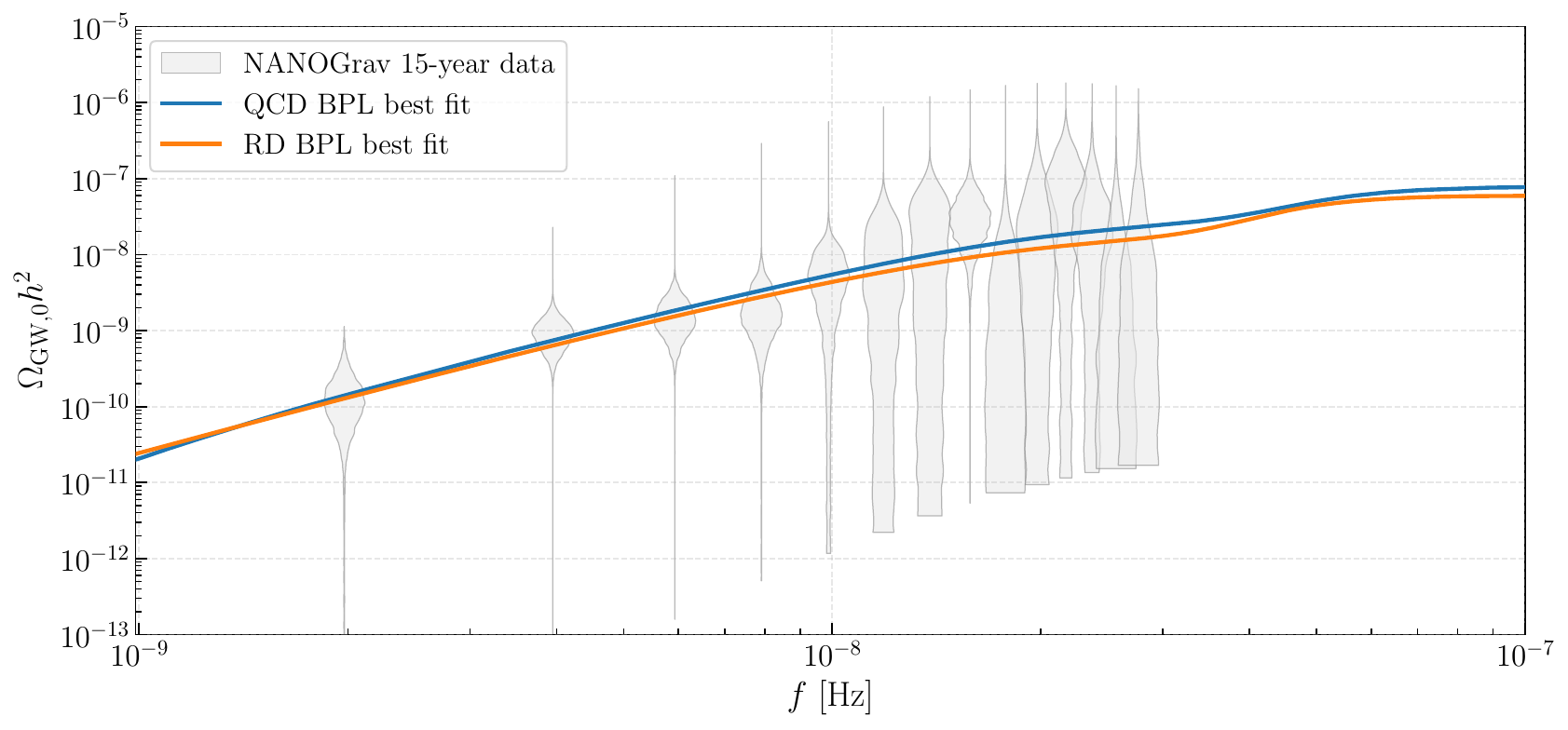}
    \caption{\textit{Top}: Best-fit delta-function spectra in the PTA frequency range, compared with the NANOGrav 15-year data. \textit{Bottom}: Best-fit broken power law spectra in the PTA frequency range, compared with the NANOGrav 15-year data.}
    \label{fig:NANOGrav_bestfit_draft}
\end{figure}

We show the GW spectra for the best-fit parameters of two choices of curvature spectra in Fig.~\ref{fig:NANOGrav_bestfit_draft}. For the delta-function case, the best fit lies at a relatively high frequency, where the relevant
modes enter well above the QCD scale. The QCD correction to the GW spectrum in the PTA band is small and thus
the Bayes factor is close to unity. The best-fit peak frequency shifts to a lower value than the radiation case, together with a lower amplitude. Although current NANOGrav data cannot distinguish best-fit spectra with and without the QCD correction, future experiments in the $\mu\textrm{Hz}$ band, such as $\mu$Ares and asteroid-based detectors, have the potential to disentangle them. 
The BPL case prefers a lower turnover
frequency, so its modes enter closer to the QCD crossover and the posterior is more visibly
modified in the PTA window. Unlike the delta-function case, the best-fit turnover frequency $k_0$ shifts to a higher value than the radiation case, accompanied by a higher amplitude. The QCD correction reshapes the spectrum near the turnover, while the two
best-fit spectra remain largely overlapping in the PTA band. The Bayes factor
\({\cal B}_{\rm QCD/RD} = 1.36\) gives a weak preference for the QCD background on the Jeffreys scale~\cite{Jeffreys}.

\section{Conclusions}
\label{sec:concl}

The standard SIGW prediction assumes the early Universe to be in standard radiation domination. This assumption, however, does not hold in the real Universe: phase transitions drive both \(w\) and \(c_s^2\) below \(1/3\), modifying the induced gravitational waves and leaving detectable imprints in their spectrum. Such transitions can also arise at higher energy scales in models beyond the Standard Model. In this work, we revisit the Fourier-space calculation of the induced spectrum across the full Standard Model thermal history and quantify how the time-dependent background reshapes it at the different energy scales.

We point out that, in a background across several energy scales, the initial conditions should be obtained from the superhorizon solution of \(\Phi_{\bf k}\) rather than imposed as in radiation domination. The standard initial condition may otherwise overestimate the amplitude and imprint an artificial bump around a certain scale. To isolate the effect of a single phase transition, we first use a toy model. Its spectrum carries a step-like feature mainly due to entropy dilution, together with an infrared peak and an ultraviolet dip. We then turn to the Standard Model background. There is a potential concern about the entropy non-conserving QCD inputs in the literature because the SIGW calculation is based on an adiabatic fluid with entropy conservation. We show that an entropy-conserving QCD background can be constructed from lattice data within the quoted error bars. The resulting SIGW spectrum shows features similar to the toy model for each individual phase transition, although the transitions partially overlap. We further find that transitions at different energy scales are probed by different gravitational wave experiments. The QCD crossover falls in the pulsar timing band, the electroweak crossover in the microhertz band, and higher-scale transitions beyond the Standard Model in the millihertz bands of LISA, Taiji and TianQin. Finally, using the NANOGrav 15-year data, we perform a Bayesian inference of the QCD correction. We obtain the best-fit spectra for the delta-function and BPL cases and the corresponding constraints. The broken power law gives a mild preference for the QCD background, with \(\mathcal{B}_{\rm QCD/RD}=1.36\). The constraints are also slightly modified: the low-frequency limits from the BPL case weaken mildly for the QCD background.

As known in the literature, the best-fit parameters of SIGWs using current PTA data typically face the issue of over-producing primordial black holes (PBH)~\cite{NANOGrav:2023hvm}. One possible solution is to incorporate non-Gaussianity in the curvature spectrum~\cite{Franciolini:2023pbf,Wang:2023ost,Inomata:2023drn,Liu:2023ymk,Inui:2024fgk,Firouzjahi:2023xke,Cai:2018dig}. Moreover, the change of $w$ and $c_s^2$ will also affect the rate of PBH formation during the QCD crossover~\cite{Carr:2019kxo,Byrnes:2018clq,Musco:2023dak,Escriva:2022bwe}. The current work provides a clear starting point for lattice simulations of SIGW production on the Standard Model background with general non-Gaussianities~\cite{Zeng:2025cer,Zeng:2025tno,Caravano:2026hca}, as well as PBH formation in a time-dependent background. We leave these studies to future work.

\section*{Acknowledgement}
We thank Soubhik Kumar for valuable discussions and feedback on a
draft of this manuscript. This work is supported by the
National Natural Science Foundation of China (Grants No. T2388102 and 12575111).

\appendix
\section{Numerical Computation Details}
\label{app:numerical}

We solve the scalar and tensor equations for the SIGW spectrum in a general time-dependent background, using the dimensionless variable
\(x=k\eta\). This choice makes the oscillation scale of each mode of order unity
and permits the same time grid to be used for different wavenumbers. Unless stated
otherwise, the integration interval is
\begin{equation}
    10^{-3}\leq x\leq 10^{3},
    \qquad N_x=8000,
    \qquad x_i\ \text{logarithmically spaced}.
    \label{eq:numerical_x_grid}
\end{equation}
The scalar equation is
integrated with an adaptive eighth-order Dormand--Prince method, with relative and
absolute tolerances \(3\times10^{-8}\) and \(10^{-10}\), respectively. The time-dependent initial condition is obtained from the superhorizon solution described in Sec.~\ref{subsec:initial condition}.
The scalar transfer function is tabulated over the range
\(k\simeq10^{-2}\)--\(10^{16}\,\mathrm{Mpc}^{-1}\), using one hundred
points per decade.
The grid is uniformly spaced in \(\log k\). The wavenumber ranges quoted in this appendix refer to the QCD (SM) thermal
history. For the toy model and the composite-Higgs case, the ranges are different but the spacing is the same. After solving the \(\Phi\) evolution numerically at each \(k\), we store \(\Phi_k(x)\) and \(\partial_\eta\Phi_k(x)\) in a cache. 

The tensor Green's function is constructed from the rescaled tensor mode \(g_k=a h_k\). The homogeneous tensor equation is
integrated in \(x=k\eta\) with the adaptive eighth-order Dormand--Prince method, with relative and absolute tolerances \(10^{-8}\) and \(10^{-10}\),
respectively, for the QCD calculation; the toy model and composite-Higgs
calculations use relative tolerance \(3\times10^{-8}\) and the same absolute tolerance.
At \(x_{\rm ini}=10^{-3}\), the two independent basis solutions are initialized by
\((g_{k,1},\partial_x g_{k,1})=(0,1)\) and
\((g_{k,2},\partial_x g_{k,2})=(1,0)\). We store the two homogeneous basis solutions \(g_{k,1}\) and \(g_{k,2}\), together with their derivatives with respect to conformal time, on the same \(x=k\eta\) grid. The retarded Green's function is assembled from them using their Wronskian. The cache stores the small-\(g\) basis solutions and their derivatives, rather than the full two-time Green's function. We tabulate these basis-solution arrays over
\(k\simeq10^1\)--\(10^{12}\,\mathrm{Mpc}^{-1}\), again using one
hundred points per decade and logarithmic spacing.

We also compute kernel values \({\cal K}=(k/{\cal H})^2 I^2(k,k_1,k_2,\eta_*)\) on the grids and store them in a cache for computing the GW spectrum. We interpolate the \(\Phi\) cache to obtain the scalar modes \(k_1\) and \(k_2\), and the Green's function cache to obtain the tensor mode \(k\). For convenience, we use the dimensionless momentum variables \(s\) and \(t\),
\begin{equation}
    s\equiv\frac{u+v-1}{\sqrt{2}},\qquad
    t\equiv\frac{u-v}{\sqrt{2}},\qquad
    u=\frac12+\frac{s+t}{\sqrt{2}},\qquad
    v=\frac12+\frac{s-t}{\sqrt{2}},
\label{eq:numerical_st_variables}
\end{equation}
where \(u=k_1/k\) and \(v=k_2/k\). The momentum conservation gives
\(|1-v|\leq u\leq1+v\) and \(s>0\), \(|t|<1/\sqrt{2}\).
The symmetry under
\(u\leftrightarrow v\) allows us to integrate over \(t\geq0\) and multiply by two.
The Jacobian is \(\mathrm du\,\mathrm dv=\mathrm ds\,\mathrm dt
=s\ln 10\,\mathrm d\log_{10}s\,\mathrm dt\).
We then tabulate the kernel for each tensor wavenumber in \(t\in[0,0.7]\) with 20
points, slightly inside the kinematic boundary \(1/\sqrt{2}\), and in
\(\log_{10}s\in [-2,2]\) with 539 points.
The grid of \(\log_{10}s\) is refined around the resonance area. Tensor wavenumbers \(k\) are
sampled from \(10^1\) to \(10^{12}\,\mathrm{Mpc}^{-1}\), with a spacing of 0.02 dex in the QCD region, \(10^{5.4}\)--\(10^{7}\,\mathrm{Mpc}^{-1}\), and 0.05 dex elsewhere.
For a fixed \((k,s,t)\) grid, we integrate the kernel from
\(\eta_{\rm ini}=10^{-3}/[k\min(1,u,v)]\) to \(\eta_*\). The
integral uses a logarithmic grid while the modes are superhorizon and a linear grid
after the onset of oscillations. We average the
kernel over four periods using 33 equally spaced phase samples at \(\eta_*={1000}/[{k\,\max(1,u,v)}]\), where the averaged kernel value has reached a constant.

Finally, we interpolate the kernel cache to compute the GW spectrum. 
\begin{align}
    \Omega_{\rm GW,0} h^2
    =&\, \Omega_{r,0} h^2\frac{4\ln 10}{3}
    \int_{-2}^{2} \d \log_{10} s\, \int_0^{1/\sqrt{2}} \d t \,s\,
    \l[\frac{2s(s+\sqrt{2})(1-2t^2)}{(1+\sqrt{2}s)^2-2t^2}\r]^2
    \nonumber\\
    &\times{\cal P}_\zeta \l(ku(s,t)\r)\,{\cal P}_\zeta \l(kv(s,t)\r)\,
    {\cal K}_{\rm cache}(k,s,t)\,R(\eta_*),
\end{align}
where \(R\) is the entropy dilution factor
\begin{equation}
    R(\eta_*)=\frac{g_{*}(\eta_*)}{g_{*}(\eta_{0})}
    \left(\frac{g_{*s}(\eta_{0})}{g_{*s}(\eta_*)}\right)^{4/3} . \label{eq:numerical_redshift_factor}
\end{equation}
In practice we evaluate \(R\) from the background tables as the equivalent ratio of \(a\mathcal H=a^2H\) between \(\eta_*\) and a late time \(\eta_{\rm ref}\) chosen after all
relevant thermal transitions, since \((a\mathcal H)^2\propto g_*g_{*s}^{-4/3}\).

The codes for generating these caches and computing GW spectrum are
available at \url{https://github.com/Eulara/Time-dependent-background-SIGW}.

\section{Time-dependent Initial Condition for \texorpdfstring{\(\Phi_k\)}{Phi k}}
\label{app:superhorizon}

This appendix derives the regular scalar mode for a spatially flat, adiabatic
single-fluid background satisfying Eq.~\eqref{eq:adiabatic_cs_w}, with negligible
anisotropic stress. On superhorizon scales, \(k/\mathcal H\ll1\), the gradient
term in Eq.~\eqref{eq:scalar_eom} is negligible and the Bardeen equation becomes
\begin{equation}
    \Phi_k''+3\mathcal H(1+c_s^2)\Phi_k'
    +3\mathcal H^2(c_s^2-w)\Phi_k=0.
    \label{eq:Phi_superhorizon}
\end{equation}
To eliminate \(w\) and \(c_s^2\) explicitly, it is convenient to use cosmic time \(t\).
With \(\mathcal H=aH\), a dot denoting \(\mathrm d/\mathrm dt\), and
\begin{equation}
    \Phi_k'=a\dot\Phi_k,
    \qquad
    \Phi_k''=a^2\left(\ddot\Phi_k+H\dot\Phi_k\right).
\end{equation}
Equation~\eqref{eq:Phi_superhorizon} becomes
\begin{equation}
    \ddot\Phi_k+H(4+3c_s^2)\dot\Phi_k
    +3H^2(c_s^2-w)\Phi_k=0.
    \label{eq:Phi_superhorizon_cosmic}
\end{equation}
For a spatially flat background sourced by a single adiabatic fluid, the Friedmann and
continuity equations imply
\begin{equation}
    w=-1-\frac{2\dot H}{3H^2},
    \qquad
    c_s^2=\frac{\dot p}{\dot\rho}
    =-1-\frac{\ddot H}{3H\dot H}.
    \label{eq:w_cs_H_superhorizon}
\end{equation}
Substituting these relations into Eq.~\eqref{eq:Phi_superhorizon_cosmic} gives the desired
equation containing only the background expansion rate,
\begin{equation}
    \ddot\Phi_k
    +\left(H-\frac{\ddot H}{\dot H}\right)\dot\Phi_k
    +\left(2\dot H-\frac{H\ddot H}{\dot H}\right)\Phi_k=0 .
\label{eq:Phi_superhorizon_H_only}
\end{equation}
This form assumes \(\dot H\ne0\), as is the case throughout the radiation and QCD eras.

It is straightforward to verify that \( \Phi_k=H/a\) is a solution to Eq.~\eqref{eq:Phi_superhorizon_H_only}. Writing the
coefficient of \(\dot\Phi_k\) in Eq.~\eqref{eq:Phi_superhorizon_H_only} as \(P(t)\), one
finds \(\exp[-\int P\,\mathrm dt]\propto\dot H/a\). Reduction of order then gives a second
solution, up to an arbitrary normalization, as
\begin{align}
    \Phi_k
    &=\frac{H(t)}{a(t)}\int^t
    a(\tilde t)\frac{\dot H(\tilde t)}{H^2(\tilde t)}\,\mathrm d\tilde t \\
    &=\frac{H(t)}{a(t)}\int^t a(\tilde t)\,\mathrm d\tilde t-1,
    \label{eq:Phi_reduction_order}
\end{align}
where the identity
\(\mathrm d(a/H)/\mathrm dt=a-a\dot H/H^2\) was used in the second line. An overall sign
is arbitrary, and a change of the lower integration limit can be absorbed into the
coefficient of \(H/a\). The general superhorizon solution is therefore
\begin{equation}
    \Phi_{\bf k}(t)=A_{\bf k}\left[1-\frac{H(t)}{a(t)}
    \int^t a(\tilde t)\,\mathrm d\tilde t\right]
    +B_{\bf k}\frac{H(t)}{a(t)}.
    \label{eq:Phi_superhorizon_cosmic_solution}
\end{equation}
Since \(\mathrm dt=a\,\mathrm d\eta\) and \(H/a=\mathcal H/a^2\), this becomes
\begin{equation}
    \Phi_{\bf k}(\eta)
    =A_{\bf k}\left[1-\frac{\mathcal H(\eta)}{a^2(\eta)}
    \int_0^\eta a^2(\tilde\eta)\,\mathrm d\tilde\eta\right]
    +B_{\bf k}\frac{\mathcal H(\eta)}{a^2(\eta)}.
    \label{eq:Phi_superhorizon_general}
\end{equation}
The second term is the decaying mode. For radiation domination the bracket
in Eq.~\eqref{eq:Phi_superhorizon_general} is \(2/3\); matching to
Eq.~\eqref{eq:Phi_zeta_constant_w} therefore fixes \(A_{\bf k}=\zeta_{\bf k}\). This also
shows directly that the regular solution satisfies the conserved-curvature relation
Eq.~\eqref{eq:zeta_Phi_superhorizon}. Dropping the decaying mode and defining

\begin{equation}
    J(\eta)\equiv\frac{1}{a^2(\eta)}
    \int_0^\eta a^2(\tilde\eta)\,\mathrm d\tilde\eta,
    \label{eq:J_superhorizon}
\end{equation}
the scalar transfer function and the derivative used in the time-dependent initial condition are
\begin{equation}
    \Phi_k=1-\mathcal HJ,
    \qquad
    \Phi_k'=-\mathcal H'J-\mathcal H+2\mathcal H^2J
    =-\mathcal H-(\mathcal H'-2\mathcal H^2)J.
    \label{eq:Phi_superhorizon_initial_data}
\end{equation}
For standard radiation domination, \(a\propto\eta\), \(\mathcal H=1/\eta\), and \(J=\eta/3\).
Equations~\eqref{eq:Phi_superhorizon_initial_data} then give
\(\Phi_k=2/3\) and \(\Phi_k'=0\). In a time-dependent background the integral retains
the earlier expansion history, so neither quantity should be reset to its radiation value
at a finite starting time. We evaluate Eq.~\eqref{eq:Phi_superhorizon_initial_data} at
\(k\eta_{\rm ini}=10^{-3}\) and use the resulting pair \((\Phi_k,\Phi_k')\) to initialize
the full equation. For consistent adiabatic input, the omitted gradient correction
is of order \((k/\mathcal H)^2\); changing the starting time within the superhorizon
regime should leave the same regular mode up to this correction and numerical errors.

\begin{figure}[h]
    \centering
    \includegraphics[width=0.48\linewidth]{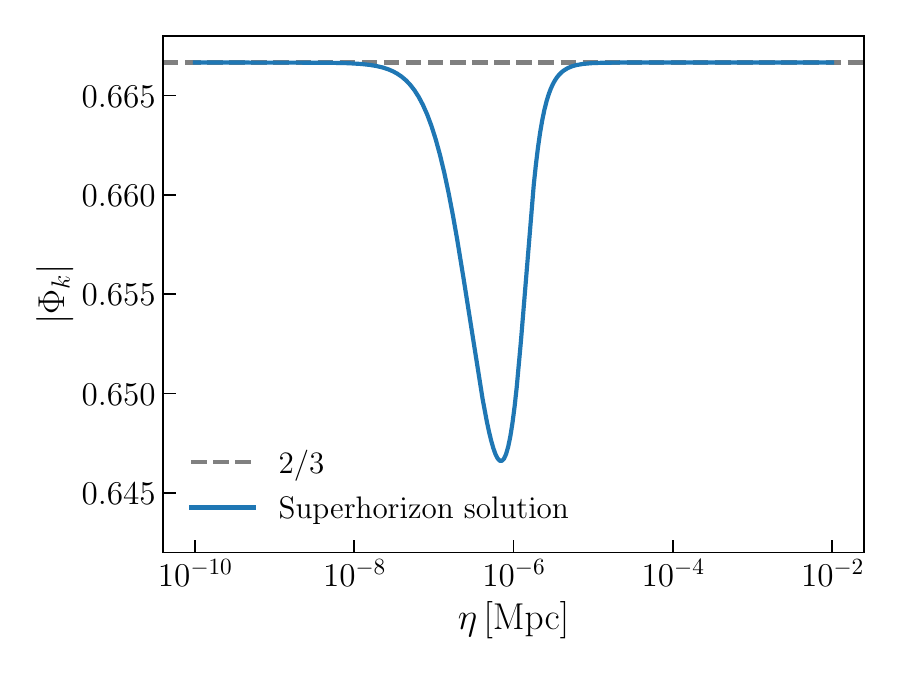}
    \includegraphics[width=0.48\linewidth]{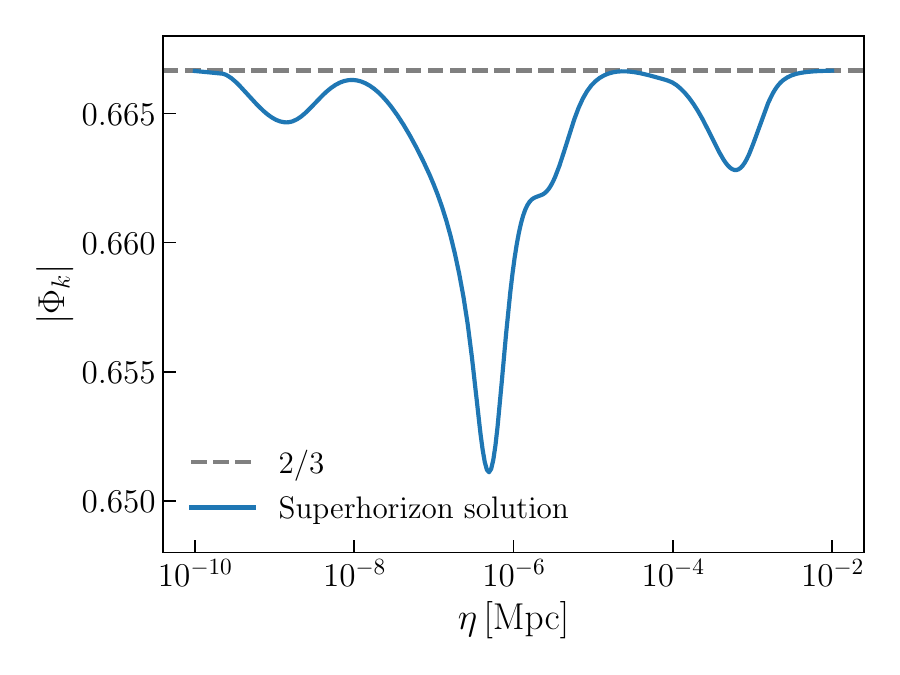}
    \caption{Superhorizon solution of the metric perturbation $\Phi_k(\eta)$ for the toy model (left) with ($r_g=10\,,m=800\,\textrm{MeV}$) and the Standard Model thermal history (right). The solid curve is the superhorizon solution of Eq.~\eqref{eq:Phi_super_conformal_draft}, and the dashed line marks \(2/3\).}
    \label{fig:superhorizon solution}
\end{figure}

\begin{figure}[h]
    \centering
    \includegraphics[width=0.48\linewidth]{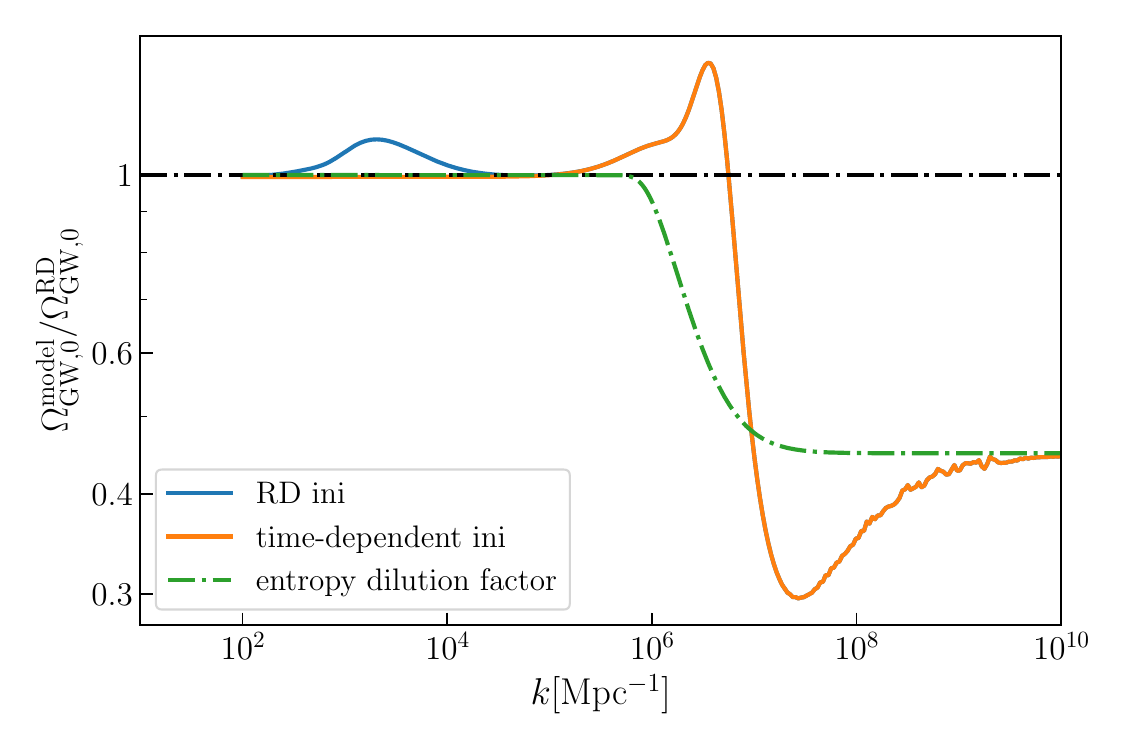}
    \includegraphics[width=0.48\linewidth]{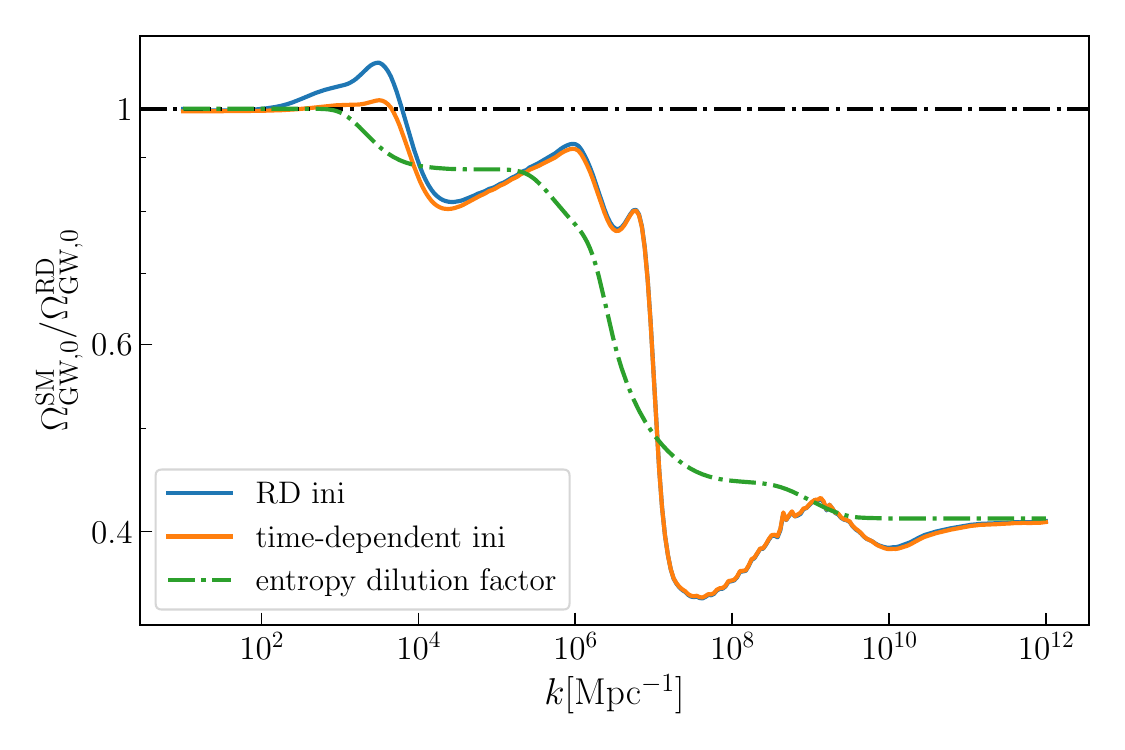}
    \caption{SIGW spectrum with a flat curvature spectrum ($\mathcal P_\zeta=A$) for the toy model (left) and the Standard Model thermal history (right), normalized to the standard radiation case. The blue and orange solid curves use the standard radiation and the time-dependent initial conditions, respectively. For both cases, we fix $k\eta_{\rm ini}=10^{-3}$. The green dot-dashed curve shows the entropy dilution factor in each case.}
    \label{fig:spectrum initial condition}
\end{figure}

As discussed in Sec.~\ref{subsec:initial condition} and shown in Fig.~\ref{fig:initial condition}, using standard radiation initial conditions with fixed $k\eta_{\rm ini}$ might lead to an artificial enhancement of the metric perturbation $\Phi_k$ for modes where $\eta_{\rm ini}$ falls in the period of the phase transition.
In this appendix, we show the superhorizon solution of Eq.~\eqref{eq:Phi_super_conformal_draft} for the toy model (left) and the Standard Model (right) in Fig.~\ref{fig:superhorizon solution}. During a transition \(\Phi_k\) decreases below its radiation value, and after the transition its superhorizon value returns to \(2/3\). This follows from the conservation of the superhorizon curvature: once the background returns to \(w=c_s^2=1/3\), \(\zeta_{\bf k}\) again fixes \(\Phi_k=2/3\), as in Eq.~\eqref{eq:Phi_zeta_constant_w}.

Figure~\ref{fig:spectrum initial condition} compares the induced GW spectrum with a flat curvature spectrum obtained with the two initial conditions, for the toy model (left) and the Standard Model (right). Away from the transition region the two spectra coincide. The difference is clearest around \(k\sim10^3\)--\(10^4\,\mathrm{Mpc}^{-1}\), where the curve that uses the standard radiation initial condition develops an artificial peak. For these modes, the initial time \(k\eta_{\rm ini}=10^{-3}\) falls in a phase transition (Sec.~\ref{subsec:initial condition}), so the radiation initial condition overestimates the amplitude at later times. This result justifies the importance of using the correct time-dependent initial conditions for metric perturbations in SIGW analysis.

\section{Details of the Toy Model}
\label{app:stepped}

This appendix summarizes the thermodynamics of the toy model used in
Sec.~\ref{sec:thermal}, following the construction of Ref.~\cite{Aloni:2022}.

Consider a coupled fluid in local thermal equilibrium at temperature \(T\), containing
\(g_{*,\rm IR}\) massless degrees of freedom and \(g_{*,\rm UV}-g_{*,\rm IR}\) degrees
of freedom sharing a common mass \(m\) (fermionic degrees of freedom carry the usual
factor of \(7/8\)). The interaction rate within the fluid is assumed to exceed the
Hubble rate, so that the fluid passes through the mass threshold in local equilibrium and
its comoving entropy is conserved exactly. Factoring out the energy density
\(\rho_B=3p_B=\pi^2T^4/30\) of a single relativistic boson degree of freedom, the
contribution of the massive particles is encoded in the dimensionless integrals
\(\hat\rho(x)\) and \(\hat p(x)\) defined as 
\begin{equation}
    \hat\rho(x)=\frac{1}{6}\int_0^\infty \mathrm{d}u\,u^2\sqrt{u^2+x^2}\,
    e^{-\sqrt{u^2+x^2}},
    \qquad
    \hat p(x)=\frac{1}{6}\int_0^\infty \mathrm{d}u\,\frac{u^4}{\sqrt{u^2+x^2}}\,
    e^{-\sqrt{u^2+x^2}},
\end{equation}
with \(x\equiv m/T\). Here the Bose and Fermi distribution functions have been
approximated by their Maxwell--Boltzmann form, which is accurate to better than \(8\%\)
for all \(x\). Both integrals equal unity at \(x\to0\) and decay exponentially for
\(x\gg1\). In terms of the ratio of massive and massless degrees of freedom \(r_g\equiv(g_{*,\rm UV}-g_{*,\rm IR})/g_{*,\rm IR}\),
the total energy density and pressure of the fluid are
\begin{equation}
    \rho(T)=g_{*,\rm IR}\rho_B(T)\left[1+r_g\hat\rho(x)\right],
    \qquad
    p(T)=g_{*,\rm IR}p_B(T)\left[1+r_g\hat p(x)\right].
    \label{eq:rho_p_stepped}
\end{equation}
The equation of state \(w=p/\rho\) is given by

\begin{equation}
    w(x)=\frac{1+r_g\hat p(x)}{3\left[1+r_g\hat\rho(x)\right]},
\end{equation}
and the sound
speed follows from \(c_s^2=\d p/\d\rho\) with the temperature dependence entering through
\(x(T)\),
\begin{equation}
    c_s^2(x)=\frac{4\left[1+r_g\hat p(x)\right]-r_g\,\d\hat p(x)/\d\ln x}
    {3\left\{4\left[1+r_g\hat\rho(x)\right]-r_g\,\d\hat\rho(x)/\d\ln x\right\}} .
    \label{eq:cs_stepped}
\end{equation}
Both quantities approach \(1/3\) for \(x\ll1\) and \(x\gg1\), and drop below it through
the transition. The dip in \(c_s^2\) is deeper and occurs slightly later than that in \(w\).

The evolution of \(x\) is fixed by exact conservation of comoving entropy,
\(s a^3=\)~const., with \(s=(\rho+p)/T\). Let \(T_{0}\) denote the temperature of the fluid today after the massive states have annihilated. Defining
\(a_t\equiv T_{0}/m\), entropy conservation gives
\begin{equation}
    \frac{x\,a_t}{a}=\left[1+\frac{r_g}{4}\left(3\hat\rho(x)+\hat p(x)\right)\right]^{1/3},
    \qquad
    a_t\equiv\frac{T_{0}}{m},
    \label{eq:entropy_stepped}
\end{equation}
which determines \(x(a)\). Together with the Friedmann equation, this relation is
equivalent to evolving \(T'/T=-3\mathcal H c_s^2\). Differentiating
\(p=w\rho\) and using \(\rho'=-3\mathcal H(1+w)\rho\) then recovers
Eq.~\eqref{eq:adiabatic_cs_w}; no independent choice of the time profiles of
\(w\) and \(c_s^2\) is involved. 
Therefore, this model is parametrized by the transition scale \(a_t\) (or equivalently $m$)
together with the ratio \(r_g\).

\section{Construction of the Entropy-conserving Standard Model Input}
\label{app:sm_eos}

\begin{figure}[ht]
    \centering
    \includegraphics[width=0.7\linewidth]{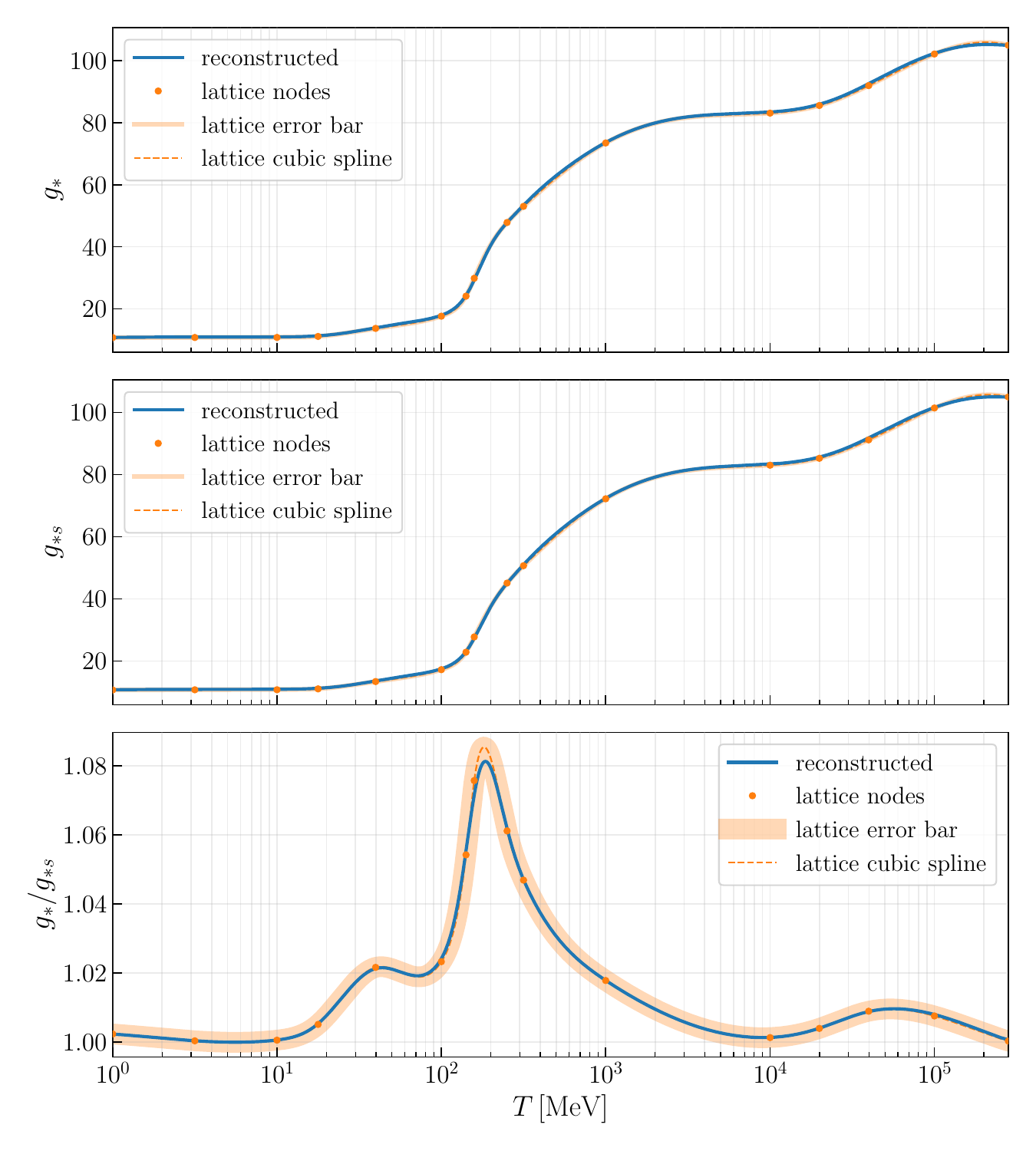}
    \caption{The reconstructed \(g_*\) (top), \(g_{*s}\) (middle), and their ratio \(g_*/g_{*s}\) (bottom) as functions of temperature, compared with the lattice QCD input of Ref.~\cite{Borsanyi:2016}. The solid curves are obtained from the entropy-conserving construction of Eqs.~\eqref{eq:app_w_from_s2}--\eqref{eq:app_reconstruct_g}; the dots mark the lattice nodes, the shaded bands their error bars, and the dash-dotted curves the cubic-spline representations of the S2 table.}
    \label{fig:app_sm_eos_reconstruction}
\end{figure}

We construct the Standard Model equation of state from the lattice-QCD
tabulation of the effective energy and entropy degrees of freedom,
denoted by \(g_*(T)\) and \(g_{*s}(T)\).
The S2 table in Ref.~\cite{Borsanyi:2016} provides \(g_*\) and the ratio \(g_*/g_{*s}\) at a set of
temperatures between \(1\,\mathrm{MeV}\) and \(10^{5.45}\,\mathrm{MeV}\), including contributions from the electroweak phase transition.
At each tabulated point we first compute
\begin{equation}
    w(T)=\frac{4}{3}\frac{g_{*s}(T)}{g_*(T)}-1
    =\frac{4}{3\,[g_*(T)/g_{*s}(T)]}-1 .
    \label{eq:app_w_from_s2}
\end{equation}
To obtain a differentiable representation, we interpolate
\(\ln g_*\) and \(\ln[g_*/g_{*s}]\) with cubic splines as functions of
\(\log_{10}(T/\mathrm{MeV})\). The resulting \(w\) is evaluated on a dense
uniform grid in \(\ln T\) and smoothed with a Gaussian kernel of width
\(\sigma_{\ln T}=0.10\). We also slightly adjust the values of $w$ in high temperature to achieve better performance. A final cubic spline of this smoothed function is
used to evaluate its derivative. The adiabatic sound speed is then fixed by
the thermodynamic relation
\begin{equation}
    c_{s,\mathrm{ad}}^2(T)=
    \frac{w(T)[1+w(T)]}
    {1+w(T)-\mathrm{d}w/\mathrm{d}\ln T} .
    \label{eq:app_cs_from_w}
\end{equation}

As a consistency check, we reconstruct the effective degrees of freedom from
the resulting adiabatic pair. Since
\(\rho=(\pi^2/30)g_* T^4\) and
\(\mathrm{d}\ln\rho/\mathrm{d}\ln T=(1+w)/c_s^2\),
\begin{equation}
    \frac{\mathrm{d}\ln g_*}{\mathrm{d}\ln T}
    =\frac{1+w}{c_s^2}-4,
    \qquad
    g_{*s}=\frac{3}{4}(1+w)g_* .
    \label{eq:app_reconstruct_g}
\end{equation}
The integration is normalized at the highest S2 table node,
\(T=10^{5.45}\,\mathrm{MeV}\), where we set
\(g_*=104.98\). We compare the reconstructed curves with cubic-spline
representations of the S2 table data. Figure~\ref{fig:app_sm_eos_reconstruction} shows that the reconstructed \(g_*\), \(g_{*s}\) and ratio \(g_*/g_{*s}\) lie entirely within the error bars of Ref.~\cite{Borsanyi:2016} (\(\Delta g_*=\Delta g_{*s}=0.75\) and \(\Delta(g_*/g_{*s})=0.003\)).

For the full Standard Model input, the reconstructed QCD curves are joined to
the low-temperature electron-positron annihilation table with a cosine blend
uniform in \(\ln T\) over \(0.7\)--\(1.4\,\mathrm{MeV}\)\cite{Musco:2023dak}. The low-temperature
end is smoothly extended to \(w=c_s^2=1/3\) between \(0.01\) and
\(0.08\,\mathrm{MeV}\), while the high-temperature end is extended to the same
radiation limit between \(2.82\times10^5\) and \(10^6\,\mathrm{MeV}\).

\section{Numerical Checks for Features in the Gravitational Wave Spectrum }
\label{sec:numerical check}

\begin{figure}[h]
    \centering
    \includegraphics[width=0.48\linewidth]{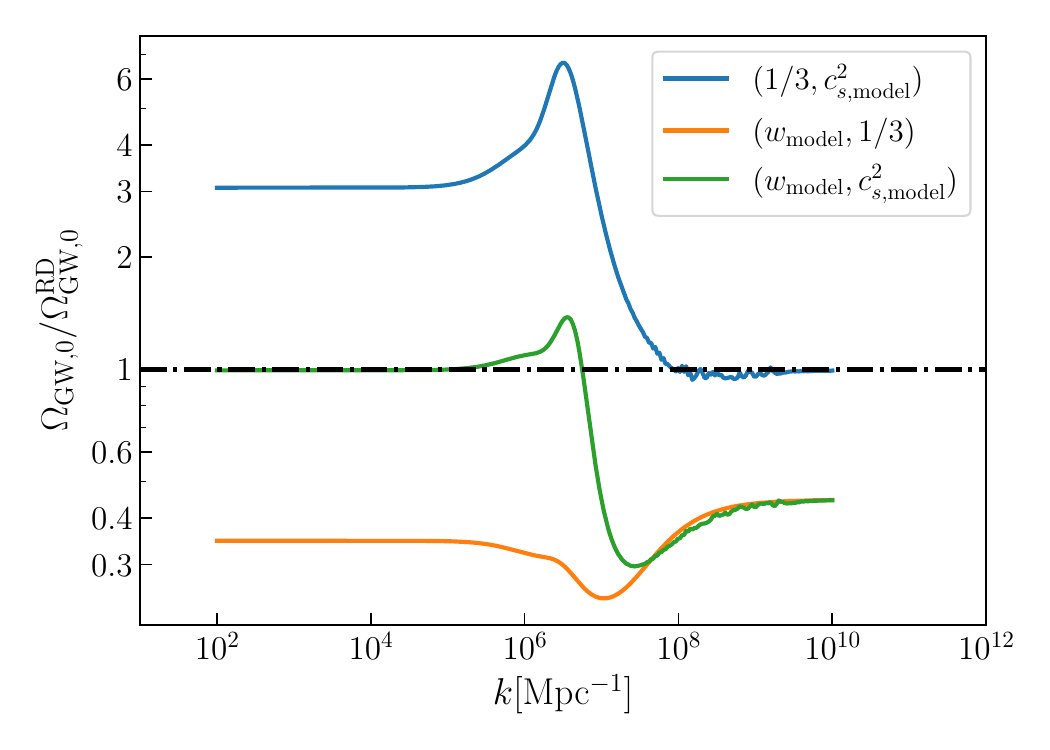}
    \includegraphics[width=0.48\linewidth]{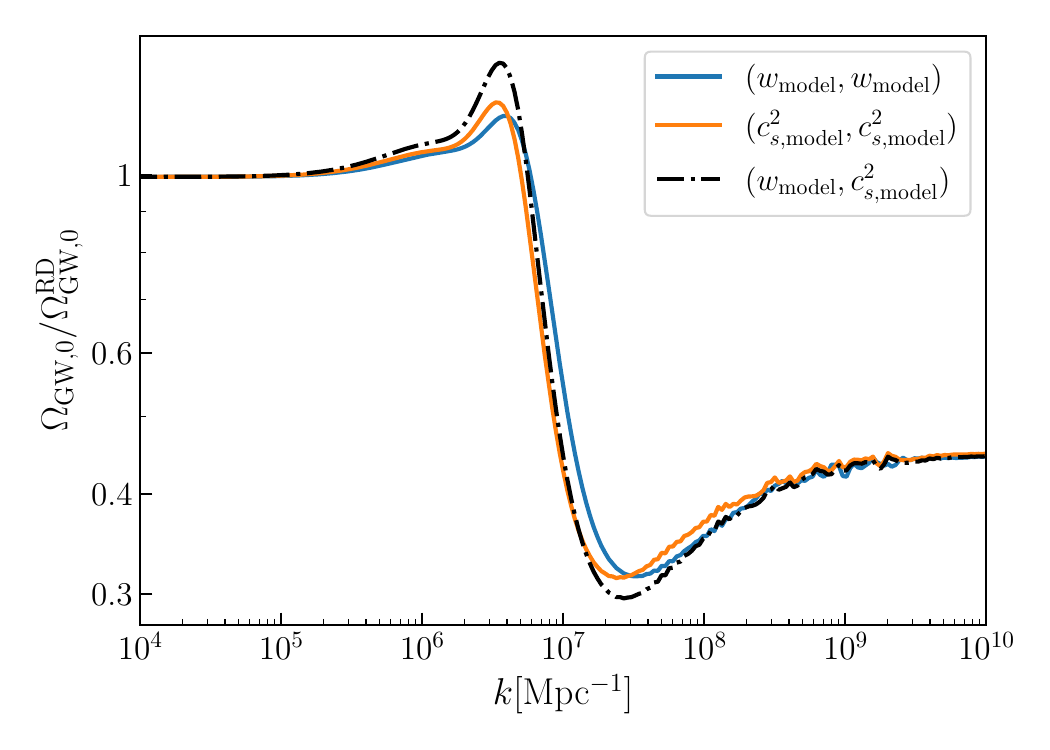}
    \caption{SIGW spectra with a scale-invariant curvature spectrum $\mathcal P_\zeta=A$ for different test parameters based on the toy model, normalized to the
    standard radiation result. The left panel shows results of fixing one of the parameters to its radiation value: $w_{\rm test}=1/3\,,c_{s,\rm test}^2=c_{s,\rm model}^2$ and $c_{s,\rm test}^2=1/3\,,w_{\rm test}=w_{\rm model}$, compared with the full toy model. The right panel shows cases with identical $w$ and $c_s^2$: $w_{\rm test}=c_{s,\rm test}^2=w_{\rm model}$ and $w_{\rm test}=c_{s,\rm test}^2=c^2_{s,\rm model}$.}
    \label{fig:w_cs_controls_draft}
\end{figure}

\begin{figure}[h]
    \centering
    \includegraphics[width=0.45\linewidth]{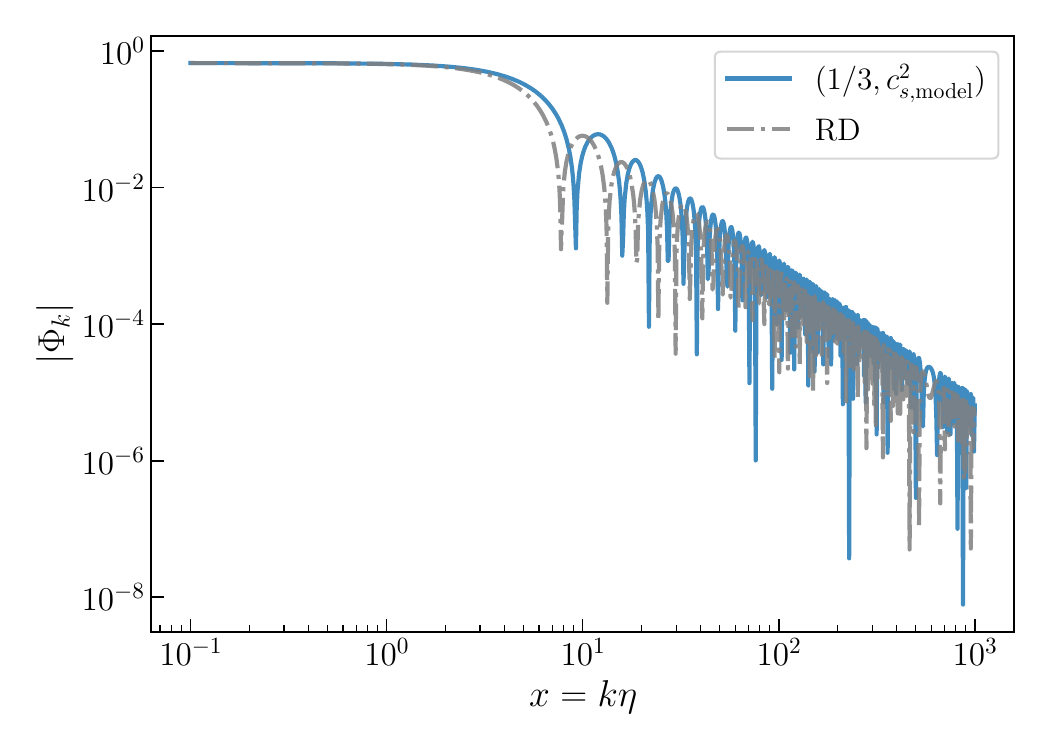}
    \includegraphics[width=0.45\linewidth]{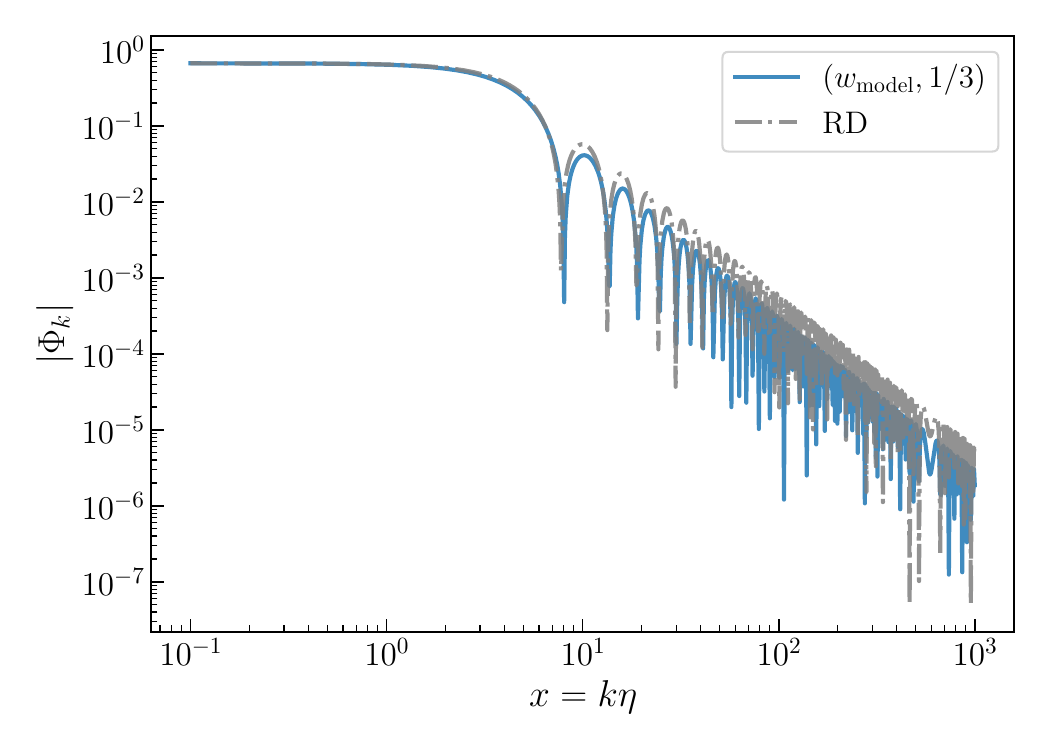}
    \caption{Scalar transfer functions with test parameters for
    \(k=1.2\times10^7\,\mathrm{Mpc}^{-1}\), compared with the standard radiation case. Left: $w_{\rm test}=1/3\,,c_{s,\rm test}^2=c_{s,\rm model}^2$. Right:
     $c_{s,\rm test}^2=1/3\,,w_{\rm test}=w_{\rm model}$.}
    \label{fig:Phi_controls_draft}
\end{figure}

As discussed in Sec.~\ref{sec:flat spectrum}, we give an intuitive argument for why the peak on the IR side is due to $c_s^2<1/3$ and the dip on the UV side is due to $w<1/3$. In this appendix, we show a few test calculations, aiming to justify previous estimates. Since both $w$ and $c_s^2$ drop below their radiation value during the transition, we manually set one of them to $1/3$ to isolate the effect of the variation of the other. Based on the toy model with parameters $(r_g=10\,,m=800\,\rm{MeV})$, we first fix \(w_{\rm test}=1/3\) and keep $c_{s,\rm test}^2(\eta)=c_{s,\rm model}^2(\eta)$, shown in the left panel of Fig.~\ref{fig:w_cs_controls_draft}. It is clear that the peak remains while the dip disappears. We then fix \(c_{s,\rm test}^2=1/3\), \(w_{\rm test}(\eta)=w_{\rm model}(\eta)\) and find that the dip remains while the peak disappears. The mismatch in the IR with the standard radiation result is due to the effective mass term (\(3\mathcal H^2(c_s^2-w)\Phi_k\) in Eq.~\eqref{eq:scalar_eom}) and entropy dilution factor for these artificial sets of parameters.
To view these effects at the level of scalar perturbations, we also show the transfer function of the mode $k=1.2\times10^7\,\textrm{Mpc}^{-1}$ around the transition scale in Fig.~\ref{fig:Phi_controls_draft}. In the left panel, the varying \(c_s^2\) leads to a delayed sound horizon entry, which enhances the amplitude relative to the standard radiation case. In the right panel, \(w<1/3\) instead suppresses the amplitude, which reduces the source and hence the GW spectrum. These numerical checks justify our previous estimate that the peak originates from the variation of \(c_s^2\), and the dip from the variation of \(w\). 

We perform an additional check of the impact of the effective mass term $3\mathcal{H}^2(c_s^2-w)\Phi_k$ in Eq.~\eqref{eq:scalar_eom}, which changes sign during the transition. We manually set $w=c_s^2$ in two test calculations such that this effective mass term vanishes: fixing $w_{\rm test}=c_{s,\rm test}^2=w_{\rm model}$ and $w_{\rm test}=c_{s,\rm test}^2=c^2_{s,\rm model}$. As seen in the right panel of Fig.~\ref{fig:w_cs_controls_draft}, both the IR peak and the UV dip survive, although weakened. This result shows that the sign change of the effective mass term does not have a significant impact on the GW spectrum. We also find that the peak position is insensitive to the value of \(w\) alone, but shifts together with \(c_s^2\). This suggests that the peak is tied to the variation of \(c_s^2\), consistent with our estimate in Sec.~\ref{sec:flat spectrum}.

\bibliographystyle{refs}
\bibliography{references}

\end{document}